\documentclass[sigplan,10pt]{acmart}
\usepackage{booktabs} 

\setcopyright{acmcopyright}

\usepackage{prooftree}
\usepackage{todonotes}

\newenvironment{changebar}{}{}
\newcommand{\changed}[1]{#1}
\newcommand{\highlight}[1]{#1}

\acmConference{European Conference on Computer Systems (EuroSys)}{February
2019}{Dresden, Germany}
\acmYear{2019}
\usepackage{amsthm}
\usepackage{scalerel}
\usepackage[utf8]{inputenc}
\usepackage[english]{babel}

\newtheorem{theorem}{Theorem}
\newtheorem{assumption}{Assumption}
\newtheorem{lemma}{Lemma}

\newcommand{\of}[1]{\left ( #1 \right )}

\newcommand{\evaluates}{\Downarrow}

\newcommand{\bnfdef}{\;\;:=\;\;}
\newcommand{\bnfsep}{\;\;|\;\;}
\newcommand{\bnfc}[1]{\operatorname{\textnormal{\textbf{#1}}}}

\newcommand{\bcjump}[1]{\bnfc{goto} \of {#1}}

\newcommand{\bccall}[3]{\bnfc{call}_{#3}\  {#1} \of{#2}}
\newcommand{\bicall}[2]{\bnfc{icall}\  {#1} \of{#2}}
\newcommand{\bcret}{\bnfc{ret}}

\newcommand{\beifthenelse}[3]{\bnfc{ifthenelse} \of {#1,#2,#3}}

\newcommand{\values}{Val}
\newcommand{\smcstate}[1]{\mem_{#1}, \regstate_{#1}}

\newcommand{\config}{s}
\newcommand{\regs}{Reg}
\newcommand{\regstate}{\rho}

\newcommand{\pc}{pc}
\newcommand{\stack}{\sigma}

\newcommand{\mem}{\mu}
\newcommand{\tuple}[1]{\langle {#1} \rangle}

\newcommand{\op}{\lozenge}
\newcommand{\unop}{\op_u}
\newcommand{\binop}{\op_b}
\newcommand{\unaryop}[1]{\unop\; #1} 
\newcommand{\binaryop}[2]{#1 \;\binop\; #2}
\newcommand{\seclabel}{\ell}

\newcommand{\hseclabel}{\mathbf{H}}
\newcommand{\lseclabel}{\mathbf{L}}

\newcommand{\seccontext}{\varGamma}

\newcommand{\funstruct}{\mathcal{F}}

\newcommand{\cfg}{G}
\newcommand{\cmagicn}{\mathit{M\_call}}
\newcommand{\rmagicn}{\mathit{M\_ret}}
\newcommand{\reg}[1]{\mathit{reg}_{#1}}

\newcommand{\secle}[2]{{#2}\sqsubseteq{#1}}
\newcommand{\sregr}{\mathit{rcaller}}
\newcommand{\srege}{\mathit{rcallee}}

\newcommand{\meth}{\mathit{f}}
\newcommand{\cntrls}{\mathit{V}}
\newcommand{\cntrl}{\mathit{v}}
\newcommand{\edges}{\mathit{E}}

\newcommand{\leqv}{=_{\lseclabel}}
\newcommand{\preimg}{\mathit{Dom}}

\newcommand{\trusted}{\mathcal{T}}
\newcommand{\untrusted}{\mathcal{U}}
\newcommand{\pred}[2]{\mathit{pred}(#1,#2)}

\newcommand{\elem}[2]{{#1}.{\mathit{#2}}}
\newcommand{\prog}{\mathcal{P}}
\newcommand{\stmt}{\mathit{Cmd}}

\newcommand{\confgiAdversary}{\lightning}
\newcommand{\mvalue}{\mathit{n}}
\newcommand{\tstate}{\nu} 
\newcommand{\tto}{\hookrightarrow}

\usepackage{hyperref}
\usepackage[normalem]{ulem}

\usepackage{afterpage}
\usepackage{color}
\usepackage{amssymb,amsmath,amsfonts}
\usepackage{extarrows}
\usepackage{version}
\usepackage{xspace}
\usepackage{graphicx}
\usepackage{listings}
\usepackage{wrapfig}
\usepackage{ifpdf}
\usepackage{stmaryrd}
\usepackage{multirow}
\usepackage{proof}
\usepackage{enumitem}
\usepackage{marginnote}
\usepackage{todonotes}
\usepackage{xspace}
\usepackage{relsize}
\usepackage{subfig}
\usepackage[font={bf,sc,small},singlelinecheck=off]{caption}

\newcommand{\comm}[3]{{\color{#1}{[#2: #3]}}}

\newcommand{\deepak}[1]{\comm{red}{Deepak}{#1}}

\def\Snospace~{\S{}}

\definecolor{mygreen}{rgb}{0,0.6,0} 
\definecolor{mygray}{rgb}{0.5,0.5,0.5}
\definecolor{mymauve}{rgb}{0.58,0,0.82}

\let\ls\lstinline

\newcommand{\private}{\ls{private}\xspace}
\newcommand{\cc}{\textsc{ConfLLVM}\xspace}  
\newcommand{\Verifier}{\textsc{ConfVerify}\xspace}
\newcommand{\ldaploc}{300,000\xspace}
\newcommand{\ldapedits}{100\xspace}
\newcommand{\ldapadds}{52\xspace}
\newcommand{\ldapfiles}{728\xspace}

\renewcommand{\U}{{\mathcal{U}}\xspace}
\newcommand{\T}{{\mathcal{T}}\xspace}
\renewcommand{\L}{{\mathcal{L}}\xspace}
\newcommand{\Mcall}{\texttt{M}_\texttt{Call}}
\newcommand{\Mret}{\texttt{M}_\texttt{Ret}}

\newcommand{\cBase}{\textbf{Base}}
\newcommand{\cCC}{$\mbox{\textbf{Our}}_{\mbox{\scriptsize Bare}}$}
\newcommand{\cCCmpx}{$\mbox{\textbf{Our}}_{\mbox{\scriptsize MPX}}$}
\newcommand{\cCCseg}{$\mbox{\textbf{Our}}_{\mbox{\scriptsize Seg}}$}
\newcommand{\cCCcfi}{$\mbox{\textbf{Our}}_{\mbox{\scriptsize CFI}}$}

\newcommand{\cBaseOA}{$\mbox{\textbf{Base}}_{\mbox{\scriptsize OA}}$}

\newcommand{\cCCnolib}{$\mbox{\textbf{Our}}_{\mbox{\scriptsize 1Mem}}$}
\newcommand{\cCConlylib}{$\mbox{\textbf{Our}}_{\mbox{\scriptsize Bare}}$}
\newcommand{\cCConlycfi}{$\mbox{\textbf{Our}}_{\mbox{\scriptsize CFI}}$}
\newcommand{\cCCsamestack}{$\mbox{\textbf{Our}}_{\mbox{\scriptsize MPX-Sep}}$}
\newcommand{\cCCfull}{$\mbox{\textbf{Our}}_{\mbox{\scriptsize MPX}}$}

\begin{document}

\copyrightyear{2019} 
\acmYear{2019} 
\setcopyright{acmlicensed}
\acmConference[EuroSys '19]{Fourteenth EuroSys Conference 2019}{March 25--28, 2019}{Dresden, Germany}
\acmBooktitle{Fourteenth EuroSys Conference 2019 (EuroSys '19), March 25--28, 2019, Dresden, Germany}
\acmPrice{15.00}
\acmDOI{10.1145/3302424.3303952}
\acmISBN{978-1-4503-6281-8/19/03}

\title{{\cc}: A Compiler for Enforcing Data Confidentiality in Low-Level Code}

\author{Ajay Brahmakshatriya}
\affiliation{
  \institution{MIT, USA}            
}
\email{ajaybr@mit.edu}
\authornote{Work done while the author was at Microsoft Research, India.}

\author{Piyus Kedia}
\affiliation{
  \institution{IIIT Delhi, India}           
}
\email{piyus@iiitd.ac.in}
\authornotemark[1]

\author{Derrick P. McKee}
\affiliation{
  \institution{Purdue University, USA}           
}
\email{mckee15@purdue.edu}
\authornotemark[1]

\author{Deepak Garg}
\affiliation{
  \institution{MPI-SWS, Germany}           
}
\email{dg@mpi-sws.org}

\author{Akash Lal}
\affiliation{
  \institution{Microsoft Research, India}           
}
\email{akashl@microsoft.com}

\author{Aseem Rastogi}
\affiliation{
  \institution{Microsoft Research, India}           
}
\email{aseemr@microsoft.com}

\author{Hamed Nemati}
\affiliation{
  \institution{CISPA, Saarland University, Germany}           
}
\email{hnnemati@kth.se}

\author{Anmol Panda}
\affiliation{
  \institution{Microsoft Research, India}           
}
\email{t-anpand@microsoft.com}

\author{Pratik Bhatu}
\affiliation{
  \institution{AMD, India}           
}
\email{pbhatu@amd.com}
\authornotemark[1]

\begin{abstract}

We present a compiler-based scheme to protect the confidentiality of
sensitive data in low-level applications (e.g.\ those written in
C) in the presence of an active adversary. In our scheme, the
programmer marks sensitive data by lightweight annotations on
the top-level definitions in the source code. The compiler then uses a
combination of static dataflow analysis, runtime instrumentation, and a
novel \emph{taint-aware} form of control-flow integrity to prevent
data leaks even in the presence of low-level attacks. To reduce
runtime overheads, the compiler uses a novel memory layout.

We implement our scheme within the LLVM framework and evaluate it on
the standard SPEC-CPU benchmarks, and on larger, real-world
applications, including the NGINX webserver and the OpenLDAP directory
server. We find that the performance overheads introduced by our
instrumentation are moderate (average 12\% on SPEC), and the
programmer effort to port the applications is minimal.

\end{abstract}

\settopmatter{printfolios=true}
\maketitle



\section{Introduction} \label{Se:Introduction}

Many programs compute on private data: Web servers use private keys
and serve private files, medical software processes private medical
records, and many machine learning models are trained on private
inputs. Bugs or exploited vulnerabilities in these programs may leak
the private data to public channels that are visible to unintended
recipients. For example, the OpenSSL buffer-overflow vulnerability
Heartbleed~\cite{Heartbleed} can be exploited to exfiltrate a web
server's private keys to the public network in cleartext. Generally
speaking, the problem here is one of \emph{information flow
  control}~\cite{Denning:1976:LMS:360051.360056}: We would like to
enforce that private data, as well as data derived from it, is never
sent out on public channels unless it has been intentionally
declassified by the program.

The standard solution to this problem is to use static dataflow
analysis or runtime taints to track how private data flows through the
program. While these methods work well in high-level, memory-safe
languages such as Java~\cite{jflow,fabric} and ML~\cite{flowcaml}, the
problem is very challenging and remains broadly open for low-level
compiled languages like C that are not memory-safe. First, the cost of
tracking taint at runtime is prohibitively high for these languages (see
Section~\ref{sec:related}). Second, static dataflow analysis cannot
guarantee data confidentiality because the lack of memory safety allows
for buffer overflow and control-flow hijack
attacks~\cite{StackSmashing, CVE-2012-0769, Mitre-Top25,
  Shacham:2007:GIF:1315245.1315313, Roemer:2012:RPS:2133375.2133377,
  Chrome-hacked}, both of which may induce data flows that cannot be
anticipated during a static analysis.

One possible approach is to start from a safe dialect of C
(e.g.\ CCured \cite{DBLP:journals/toplas/NeculaCHMW05}, Deputy
\cite{Condit:2007:DTL:1762174.1762221}, or SoftBound
\cite{DBLP:conf/pldi/NagarakatteZMZ09}) and leverage existing
approaches such as information-flow type systems for type-safe
languages \cite{Heintze:1998:SCP:268946.268976,
  Sabelfeld:2006:LIS:2312191.2314769}.  The use of safe dialects,
however, (a) requires additional annotations and program restructuring
that cause significant programming
overhead~\cite{DBLP:journals/toplas/NeculaCHMW05,
  Lu05bugbench:benchmarks}, (b) are not always backwards compatible
with legacy code, and (c) have prohibitive runtime overhead making
them a non-starter in practical applications~(see further discussion
in Section~\ref{sec:related}).

In this paper, we present the first end-to-end, practical
compiler-based scheme
to enforce data confidentiality in C programs even in the presence of
active, low-level attacks.
Our scheme is based on the insight that complete memory safety and perfect
control-flow integrity (CFI) are neither sufficient nor necessary for preventing
data leaks. We design a compiler that is able to guarantee data confidentiality
without requiring these properties.

We accompany our scheme with formal security proofs and a
thorough empirical evaluation to show that it can be used for real
applications. Below, we highlight the main components of
our scheme.
 

\noindent{\textbf{Annotating private data.}}  We require the
programmer to add \ls{private} type annotations only to top-level
function signatures and globals' definitions to mark private
data. The programmer is free to use all of the C language, including
pointer casts, aliasing, indirect calls, varargs, variable length
arrays, etc.



\noindent{\textbf{Flow analysis and runtime instrumentation.}}  Our
compiler performs standard dataflow analysis, statically propagating
taint from global variables and function arguments that have been
marked \ls{private} by the programmer, to detect any data
leaks. This analysis \textit{assumes} the correctness of declared taints
of each pointer. These assumptions are then protected by inserting runtime checks
that \textit{assert} correctness of the assumed taint at runtime. These checks
are necessary to guard against incorrect casts, memory errors and low-level
attacks. Our design does not require any static alias analysis, which is often
hard to get right with acceptable precision.


\noindent{\textbf{Novel memory layout.}}  To reduce the overhead associated with
runtime checks, our compiler partitions
the program's virtual address space into a contiguous public region
and a disjoint, contiguous private region, each with its own stack and
heap. Checking the taint of a pointer then simply reduces to
a range check on its value. We describe two partitioning schemes,
one based on the Intel MPX ISA extension~\cite{intel-mpx}, and the
other based on segment registers
(Section~\ref{sec:TaintEnforcement}). Our runtime checks do not
enforce full memory safety: a private (public) pointer may point
outside its expected object but our checks ensure that it always
dereferences \emph{somewhere} in the private (public) region. These
lightweight checks suffice for protecting data confidentiality.

\noindent{\textbf{Information flow-aware CFI.}}  Similar to memory
safety, we observe that perfect CFI is neither necessary nor
sufficient for information flow. We introduce an efficient,
taint-aware CFI scheme. Our compiler instruments the targets of
indirect calls with \emph{magic sequences} that summarize the output
of the static dataflow analysis at those points. At the source of the
indirect calls, a runtime check ensures that the magic sequence at the
target is taint-consistent with the output of the static dataflow
analysis at the source (Section~\ref{sec:cfi}).

\noindent{\textbf{Trusted components.}}  For selective
declassification, as required for most practical applications, we
allow the programmer to re-factor \emph{trusted} declassification
functions into a separate component, which we call $\T$. The remaining
untrusted application, in contrast, is called $\U$. Code in $\T$ is
not subject to any taint checks and can be compiled using a vanilla
compiler. It can access all of $\U$'s memory, specifically, it can
perform declassification by copying data from $\U$'s private region to
$\U$'s public region (e.g. after encrypting the data). $\T$ has its
own separate stack and heap, and we re-use the range checks on memory
accesses in $\U$ to prevent them from reading or writing to $\T$'s
memory. We give an example and general guidelines for refactoring $\U$
and $\T$ in Section~\ref{sec:threat-model}.


\noindent{\textbf{Formal guarantees.}} We have formalized our scheme
using a core language of memory instructions (load and store) and
control flow instructions (goto, conditional, and direct and indirect
function call). We prove a termination-insensitive non-interference
theorem for $\U$, assuming that the $\T$ functions it calls are
non-interfering. In other words, we prove that if two
public-equivalent configurations of $\U$ take a step each, then the
resulting configurations are also public-equivalent. Our formal model
shows the impossibility of sensitive data leaks even in the presence
of features like aliasing and casting, and low-level vulnerabilities
such as buffer overflows and ROP attacks (Appendix~\ref{appendix:A}).

\paragraph{Implementation and Evalution}
We have implemented our compiler, \cc, as well as the complementary
low-level verifier, \Verifier, within the LLVM
framework~\cite{Lattner:2004:LCF:977395.977673}. We evaluate our
implementation on the standard SPEC-CPU benchmarks and three large
applications---NGINX web server~\cite{nginx}, OpenLDAP directory server~\cite{openldap},
and a neural-network-based image classifier built on
Torch~\cite{th,thnn}. All three applications have private
data---private user files in the case of NGINX, stored user passwords in
OpenLDAP, and a model trained on private inputs in the classifier. In
all cases, we are able to enforce confidentiality for the private data
with a moderate overhead on performance and a small amount of
programmer effort for adding annotations and declassification code.

\section{Overview}
\label{sec:threat-model}

\noindent{\textbf{Threat model.}} We consider C applications that work
with both private and public
data.  Applications interact with the external world using the
network, disk, and other channels. They communicate public data in
clear, but want to protect the confidentiality of the private data by,
for example, encrypting it before sending it out. However, the
application could have logical or memory errors, or exploitable
vulnerabilities that may cause private data to be leaked out in
clear.

The attacker interacts with the application and may send
carefully-crafted inputs that trigger bugs in the application. The
attacker can also observe all the external communication of the
application. Our goal is to prevent the private data of the
application from leaking out in clear. Specifically, we address
\textit{explicit} information flow: any data directly derived
from private data is also treated as private. While this addresses
most commonly occurring exploits~\cite{taintcheck}, optionally, our
scheme can be used in a stricter mode where it disallows branching on
private data, thereby preventing implicit leaks too. We ran all our
experiments (Section~\ref{sec:evaluation}) in this stricter
mode. Side channels (such as execution time and memory-access
patterns) are outside the scope of this work.

Our scheme can also be used for integrity protection in a setting
where an application computes over trusted and untrusted
data~\cite{biba}. Any
data (explicitly) derived from untrusted inputs cannot be supplied to
a sink that expects trusted data (Section~\ref{sec:sys-libs} shows an
example).

\begin{figure*}[t]
\captionsetup{justification=centering}
\begin{tabular}{l@{\hskip 0.5cm}r}

{\begin{lstlisting}[escapechar=$,frame=None]
void handleReq (char *uname, char *upasswd, char *fname,
                char *out, int out_size)
{
  char passwd[SIZE], fcontents[SIZE];$\label{handleReq:stackbufs}$
  read_password (uname, passwd, SIZE);
  if(!(authenticate (uname, upasswd, passwd))) {
    return;
  }
  //inadvertently copying the password to the log file
  send(log_file, passwd, SIZE);$\label{handleReq:logfile}$
  
  read_file(fname, fcontents, SIZE);
  //(out_size $>$ SIZE) can leak passwd to out
  memcpy(out, fcontents, out_size);$\label{handleReq:memcpy}$
  //a bug in the fmt string can print stack contents 
  sprintf(out + SIZE, fmt, "Request complete");$\label{handleReq:printf}$
}
\end{lstlisting}}

&

{\begin{lstlisting}[escapechar=$,frame=None]
#define SIZE 512

int main (int argc, char **argv)
{
  ... //variable declarations
  while (1) {
    n = recv(fd, buf, buf_size);$\label{main:recv}$
    parse(buf, uname, upasswd_enc, fname);
    decrypt(upasswd_enc, upasswd);
    handleReq(uname, upasswd, fname, out,
              size);
    format(out, size, buf, buf_Size);
    send(fd, buf, buf_size);
  }
}
\end{lstlisting}}

\end{tabular}
\caption{Request handling code for a web server}
\label{Fi:ExampleCode}
\end{figure*}

\noindent{\textbf{Example application.}} Consider the code for a web
server in ~\autoref{Fi:ExampleCode}. The server receives requests from
the
user (\ls{main}:\ref{main:recv}), where the request contains the
username and a file name (both in clear text), and the encrypted user
password. The server decrypts the password and calls the
\ls{handleReq} helper routine that copies the (public) file
contents into the \ls{out} buffer. The server finally prepares the formatted
response (\ls{format}), and sends the response (\ls{buf}), in clear,
to the user.

The \ls{handleReq} function allocates two local buffers,
\ls{passwd} and \ls{fcontents}
(\ls{handleReq}:\ref{handleReq:stackbufs}). It reads the actual
user password (e.g., from a database) into \ls{passwd}, and
authenticates the
user. On successful authentication, it reads the file contents
into \ls{fcontents},
copies them to the \ls{out} buffer, and appends a message to it
signalling the completion of the request.

The code has several bugs that can cause it to leak the user
password. First, at line~\ref{handleReq:logfile}, the code
leaks the clear-text password to a log file by mistake. Note that memory-safety
alone would not prevent this kind of bugs.
Second, at line~\ref{handleReq:memcpy}, \ls{memcpy} reads
\ls{out_size} bytes from \ls{fcontents} and copies them to \ls{out}. If
\ls{out_size} is greater than \ls{SIZE}, this can cause \ls{passwd} to
be copied to \ls{out} because an overflow past
\ls{fcontents} would go into the \ls{passwd} buffer. Third, if the format
string \ls{fmt} in the \ls{sprintf} call (line~\ref{handleReq:printf})
contains extra formatting directives, it can
print stack contents into \ls{out}
(\cite{Shankar:2001:DFS:1251327.1251343}). The situation is worse if
\ls{out_size} or \ls{fmt} can be influenced by the
attacker.

Our goal is to prevent such vulnerabilities from leaking out
sensitive application data. Below we discuss the three main components of our
approach.

\begin{figure}
  \captionsetup{justification=centering}
  \includegraphics[width=1.0\linewidth]{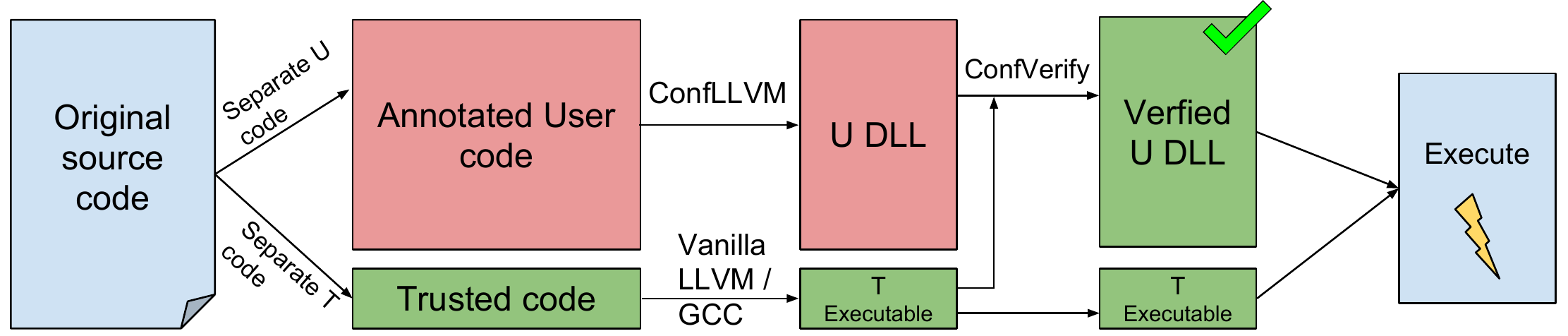}
  \caption{Workflow of our scheme and toolchain}
  \label{fig:workflow}
\end{figure}

\noindent\textbf{Identifying trusted code.}
Figure~\ref{fig:workflow} shows the workflow of our
toolchain.
The programmer starts by identifying code that must be
\emph{trusted}. This code, called $\T$
(for trusted), consists of functions that legitimately or
intentionally declassify private data, or provide I/O. The remaining
bulk of code, denoted $\U$, is \emph{untrusted} and subjected to our
compilation scheme. A good practice is to contain most of the
application logic in $\U$ and limit $\T$ to a library of generic
routines that can be hardened over time, possibly even formally
verified~\cite{pldi16}. For example, in the web server code
from~\autoref{Fi:ExampleCode}, $\T$ would consist of: \ls{recv},
\ls{send}, \ls{read_file} (network, I/O), \ls{decrypt} (cryptographic
primitive), and \ls{read_passwd} (source of sensitive data). The
remaining web server code (\ls{parse}, \ls{format}, and even
\ls{sprintf} and \ls{memcpy}) would be in $\U$.

The programmer compiles $\T$ with any compiler (or even uses
pre-compiled binaries), but $\U$ is compiled with our compiler \cc.



\noindent\textbf{Partitioning $\U$'s memory.}  To enforce
confidentiality in $\U$, we minimally require the programmer to tell
{\cc} where private data enters and exits $\U$. Since $\U$ relies on
$\T$ for I/O and communication, the programmer does so by marking
private data
in the signatures of all functions exported from $\T$ to $\U$ with a
new type qualifier \ls{private}~\cite{foster:pldi99}. Additionally, to
help {\cc}'s analysis, the programmer must annotate private data in
$\U$'s top-level definitions, i.e., globals, function signatures, and
in \ls{struct} definitions. These latter annotations within $\U$ are
\emph{not trusted}. Getting them wrong may cause a static error or
runtime failure, but cannot leak private data. {\cc} needs no other
input from the programmer. Using a dataflow analysis
(Section~\ref{sec:compiler}), it automatically infers which local
variables carry private data.  Based on this information, \cc
partitions $\U$'s memory into two regions, one for public and one for
private data, with each region having its own stack and heap. A third
region of memory, with its own heap and stack, is reserved for $\T$'s
use.




In our example, the trusted annotated
signatures of $\T$ against which \cc compiles $\U$ are:
\begin{lstlisting}[frame=None,numbers=none]
int recv(int fd, char *buf, int buf_size);
int send(int fd, char *buf, int buf_size);
void decrypt(char *ciphertxt, private char *data);
void read_passwd(char *uname, private char *pass,
                 int size);
\end{lstlisting}
while the untrusted annotations for $\U$ are:
\begin{lstlisting}[frame=None,numbers=none]
void handleReq(char *uname, private char *upasswd,
               char *fname, char *out, int out_sz);
int authenticate(char *uname, private char *upass,
                 private char *pass);
\end{lstlisting}
\cc automatically infers that, for example, \ls{passwd}
(line~\ref{handleReq:stackbufs}) is a
\ls{private} buffer. Based on this and \ls{send}'s prototype, \cc
raises a compile-time error flagging the bug at
line~\ref{handleReq:logfile}. Once the bug is fixed by the programmer
(e.g. by removing the line), \cc compiles the program and lays out the
stack and heap data in their corresponding regions
(Section~\ref{sec:TaintEnforcement}). The remaining two bugs are
prevented by runtime checks that we briefly describe next.


\noindent\textbf{Runtime checks.}  {\cc} inserts runtime checks to
ensure that, (a) at runtime, the pointers belong to their annotated or inferred
regions (e.g. a \ls{private char *} actually points to the private
region in $\U$), (b) $\U$ does not read or write beyond its own memory (i.e.
it does not read or write to $\T$'s memory), and (c) $\U$
follows a \emph{taint-aware} form of CFI that prevents circumvention
of these checks and prevents data leaks due to control-flow attacks. In
particular, the bugs on lines~\ref{handleReq:memcpy}
and~\ref{handleReq:printf} in our example cannot be exploited due to
check (a). We describe the details of these checks in
Sections~\ref{sec:TaintEnforcement} and~\ref{sec:cfi}.

$\T$ code is allowed to access all memory. However, $\T$ functions
must check their arguments to ensure that
the data passed by $\U$ has the correct sensitivity label. For
example, the \ls{read_passwd} function would check that the range
[\ls{pass}, \ls{pass+size-1}] falls inside the private memory segment
of $\U$. (Note that this is different from complete memory safety;
$\T$ need not know the size of the \ls{passwd} buffer.)




\noindent\textbf{Trusted Computing Base (TCB).}
We have also designed and implemented a static verifier, \Verifier, to
confirm that a
binary output by \cc has enough checks in place to guarantee
confidentiality (Section~\ref{sec:compiler}). 
\begin{changebar}
\Verifier guards against bugs in the compiler. It allows us to extend the threat
model to one where the adversary has full control of the compiler that was used
to generate the binary of the untrusted code.
\end{changebar}
%
To summarize, our TCB, and thus the security of our scheme, does not
depend on the (large) untrusted application code $\U$ or the
compiler. We only trust the (small) library code $\T$ and the static
verifier. We discuss more design considerations for $\T$ in
Section~\ref{sec:limitations}.

%


\section{Memory Partitioning Schemes}
\label{sec:TaintEnforcement}

\cc uses the programmer-supplied annotations, and with the help of
type inference, statically determines the taint of each memory access
(Section~\ref{sec:compiler}), i.e., for every memory load and store in
$\U$, it infers if the address contains private or public
data. It is possible for the type-inference to detect a problem (for
instance, when a variable holding private data is passed to a method
expecting a public argument), in which case, a type error is reported
back to the programmer.  On successful inference, \cc proceeds to
compile $\U$ with a custom memory layout and runtime instrumentation.
We have developed two different memory layouts as well as runtime
instrumentation schemes. The schemes have different
trade-offs, but they share the common idea -- all private and all
public data are stored in their own respective contiguous regions of
memory and the instrumentation ensures that at runtime each pointer
respects its statically inferred taint. We describe these schemes next.

\captionsetup[subfloat]{captionskip=0.2cm}

\begin{figure}[tb]
\captionsetup{justification=centering}
\subfloat[Segment scheme] {
\includegraphics[width=.4\linewidth]{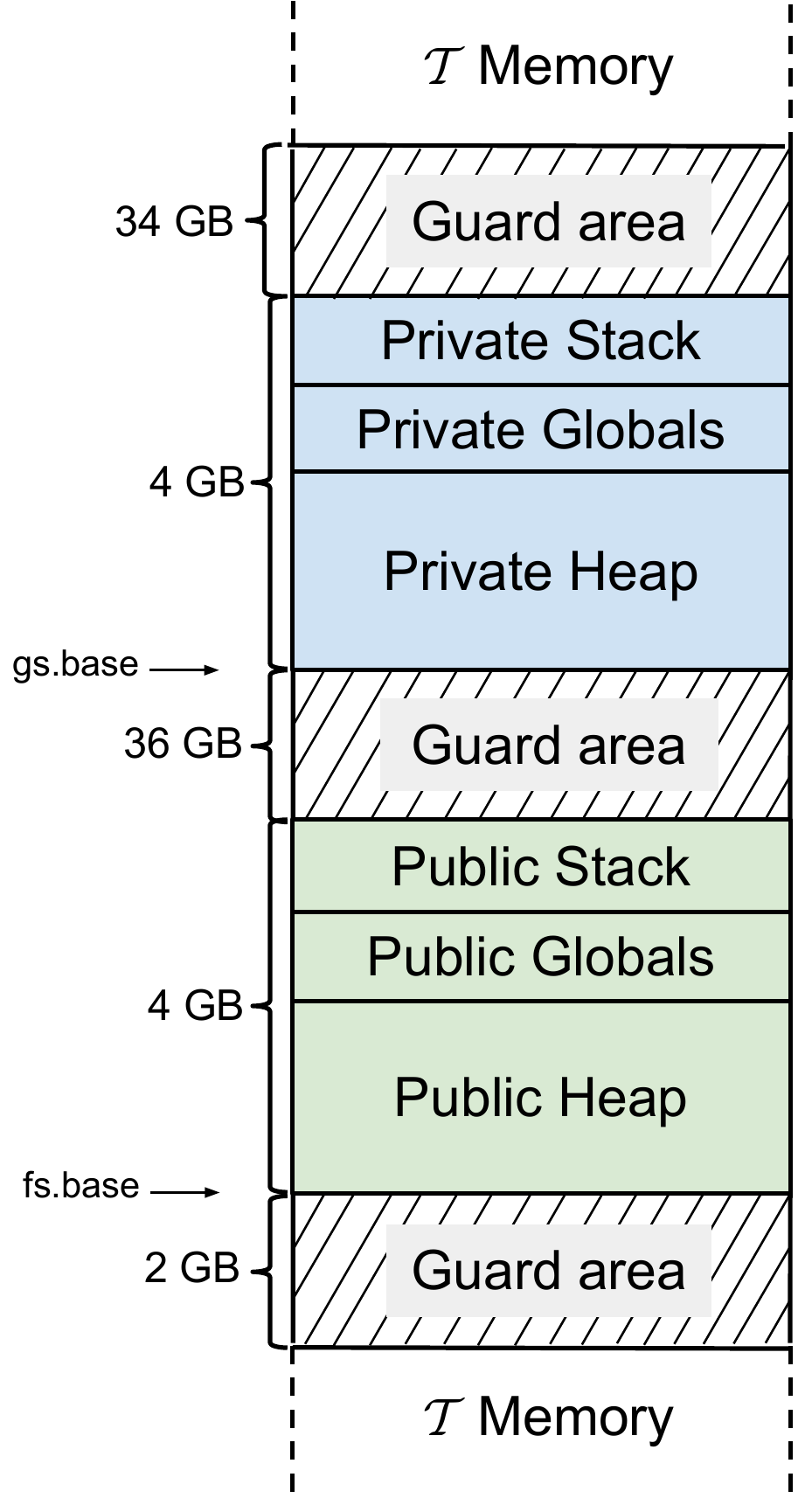}
\label{fig:memlayout-seg}}
\subfloat[MPX scheme] {
\includegraphics[width=.58\linewidth]{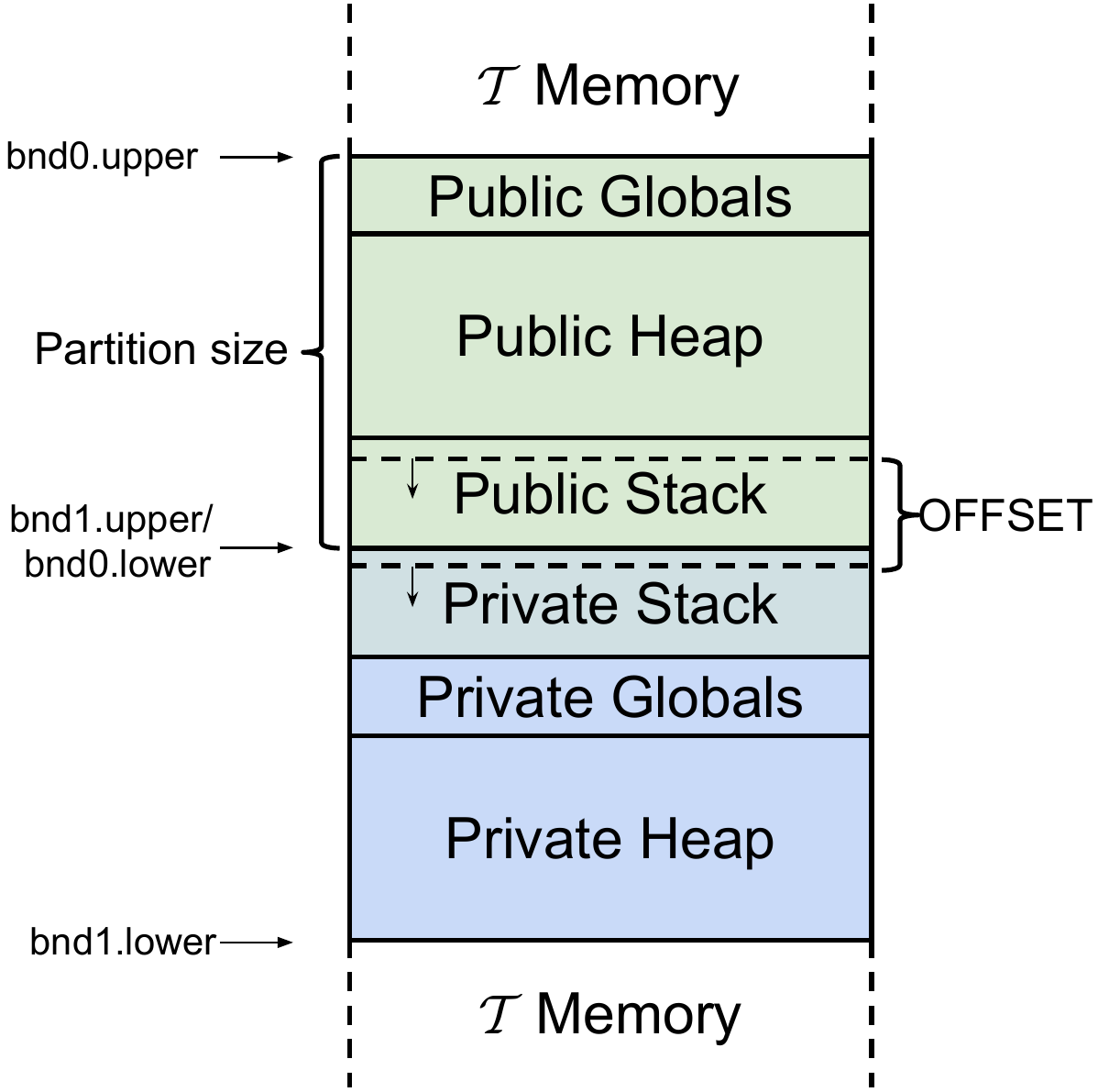}
\label{fig:memlayout-mpx}}
\caption{Memory layout of $\U$}
\label{fig:MemoryLayout}
\end{figure}

\noindent{\textbf{MPX scheme.}}  This scheme relies on the Intel MPX
ISA extension~\cite{intel-mpx} and uses the memory layout shown in
Figure~\ref{fig:memlayout-mpx}. The memory is partitioned into a
public region and a private region, each with its own heap, stack and
global segments. The two regions are laid out contiguously in
memory. Their ranges are stored in MPX bounds registers (\ls{bnd0} and
\ls{bnd1}). Each memory access is preceded with MPX instructions
(\ls{bndcu} and \ls{bndcl}) that check their first argument against
the (upper and lower) bounds of their second argument. The concrete
values to be stored in the bounds registers are determined at load
time (Section~\ref{sec:ToolChain}).

\captionsetup[subfloat]{captionskip=-0.3cm}

\newsavebox{\firstlisting}
\begin{lrbox}{\firstlisting}
\begin{lstlisting}[frame=None,numbers=none]
private int bar (private int *p, int *q)
{
  int x = *p;
  int y = *q;
  return x + y;
}
\end{lstlisting}
\end{lrbox}
\newsavebox{\secondlisting}
\begin{lrbox}{\secondlisting}
\begin{lstlisting}[language={[x86masm]Assembler},frame=None,numbers=none]
;argument registers p = r1, q = r2
;stack offsets from rsp: x: 4, y: 8
sub rsp, 16                ;rsp = rsp - 16
bndcu [r1], bnd1           ;MPX instructions to check that-
bndcl [r1], bnd1           ;-r1 points to private region
r3 = load [r1]             ;r3 = *p
bndcu [rsp+4+OFFSET], bnd1 ;check that rsp+4+OFFSET-
bndcl [rsp+4+OFFSET], bnd1 ;-points to private region
store [rsp+4+OFFSET], r3   ;x = r3
bndcu [r2], bnd0           ;check that r2 points to-
bndcl [r2], bnd0           ;-the public region
r4 = load [r2]             ;r4 = *q
bndcu [rsp+8], bnd0        ;check that rsp+8 points to-
bndcl [rsp+8], bnd0        ;-the public region
store [rsp+8], r4          ;y = r4
bndcu [rsp+4+OFFSET], bnd1
bndcl [rsp+4+OFFSET], bnd1
r5 = load [rsp+4+OFFSET]   ;r5 = x
bndcu [rsp+8], bnd0
bndcl [rsp+8], bnd0
r6 = load [rsp+8]          ;r6 = y
r7 = r5 + r6 
add rsp, 16                ;rsp = rsp + 16
ret r7
\end{lstlisting}
\end{lrbox}
\newsavebox{\thirdlisting}
\begin{lrbox}{\thirdlisting}
\begin{lstlisting}[language={[x86masm]Assembler},frame=None,numbers=none]
  ;argument registers p = r1, p = r2
  ;private memory operands are gs prefixed
  ;public  memory operands are fs prefixed

  sub rsp, 16          ;rsp = rsp - 16
  r3 = load gs:[e1]    ;r3 = *p
  store gs:[esp+4], r3 ;x = r3
  r4 = load fs:[e2]    ;r4 = *q
  store fs:[esp+8], r4 ;y = r4
  r5 = load gs:[esp+4] ;r5 = x
  r6 = load fs:[esp+8] ;r6 = y
  r7 = r5 + r6 
  add rsp, 16          ;rsp = rsp + 16
  ret r7
\end{lstlisting}
\end{lrbox}

\begin{figure*}[t]
\captionsetup{justification=centering}
\begin{minipage}{.45\textwidth}
\subfloat[A sample $\U$ procedure]{\usebox{\firstlisting}\label{fig:input}}\\
\subfloat[Assembly code under segment scheme]{\usebox{\thirdlisting} \label{fig:out-seg}}
\end{minipage}
\begin{minipage}{.52\textwidth}
\subfloat[Assembly code under MPX scheme]{\usebox{\secondlisting} \label{fig:out-mpx}}
\end{minipage}
\caption{The (unoptimized) assembly generated by \cc for an example procedure.}
\end{figure*}

The user selects the maximum stack size
\ls{OFFSET} at compile time (at most $2^{31}-1$). The scheme maintains the public and private stacks in
lock-step: their
respective top-of-stack are always at offset \ls{OFFSET} to each other.
For each function call, the compiler arranges to generate a frame each
on the public and the private stacks. Spilled \ls{private} local
variables and \ls{private} arguments are stored on the private stack;
everything else is on the public stack. Consider the procedure in
Figure~\ref{fig:input}. The generated (unoptimized) assembly under the
MPX scheme (using virtual registers for simplicity) is shown in
Figure~\ref{fig:out-mpx}.  \cc automatically infers that \ls{x} is a
\ls{private int} and places it on the private stack, whereas \ls{y} is
kept on the public stack. The stack pointer \ls{rsp} points to the top
of the public stack. Because the two stacks are kept at constant
\ls{OFFSET} to each other, \ls{x} is accessed simply as
\ls{rsp+4+OFFSET}. The code also shows the instrumented MPX bound
check instructions.

\noindent{\textbf{Segmentation scheme.}}  x64 memory operands are in
the form $[\textit{base} + \textit{index} * \textit{scale} +
  \textit{displacement}]$, where \textit{base} and \textit{index} are
$64$-bit unsigned registers, \textit{scale} is a constant with maximum
value of $8$, and \textit{displacement} is a $32$-bit signed
constant. The architecture also provides two segment registers \ls{fs}
and \ls{gs} for the \textit{base} address computation; conceptually,
\ls{fs}:\textit{base} simply adds \ls{fs} to \textit{base}.


We use these segment registers to store the
lower bounds of the public and private memory regions, respectively,
and prefix the \emph{base} of memory operands with these registers.
The public and private regions are separated 
by (at least) $36$GB of guard space (unmapped
pages that cause a fault when accessed).
The guard sizes are chosen so that any memory operand
whose \textit{base} is prefixed with \ls{fs} cannot escape the public
segment, and any memory operand prefixed with \ls{gs} cannot escape
the private segment (Figure~\ref{fig:memlayout-seg}).

The segments are each aligned to a $4$GB boundary. The usable space within each
segment is also $4$GB. We access the \emph{base} address stored in a $64$-bit register, say a private
value stored in \ls{rax}, as \ls{fs+eax}, where \ls{eax} is the 
lower 32 bits of \ls{rax}. Thus, in \ls{fs+eax}, the lower $32$ bits
come from \ls{eax} and the upper 32 bits come from \ls{fs} (because \ls{fs} is $4$GB
aligned). 
\begin{changebar}
Further, the index register is also constrained to use lower $32$ bits only. 
\end{changebar}
This implies that the maximum offset within a segment
that $\U$ can access is $38$GB ($4 + 4*8 + 2$). This is rounded up
to $40$GB for $4$GB alignment, with $4$GB of usable space and $36$GB
of guard space. Since the \textit{displacement} value can be
negative, the maximum negative offset is $2$GB, for which we have
the guard space below the public segment.

The usable parts of the segments are restricted to $4$GB because it is
the maximum addressable size using a single $32$ bit register. This
restriction also ensures that we don't have to
translate $\U$ pointers when the control is passed to $\T$, thus
avoiding the need to change
or recompile $\T$. Generated code for our example under this scheme is shown in
Figure~\ref{fig:out-seg}. The figure uses the
convention that \ls{e}$_i$ (resp., \ls{esp}) represents the lower $32$ bits of the register 
\ls{r}$_i$ (resp., \ls{rsp}). The public and private stacks are still
maintained in lock-step. Taking the address of a private
stack variable requires extra support: 
the address of variable \ls{x} in our example is \ls{rsp+4+size}, where \ls{size} is the total
segment size ($40$GB).

The segmentation scheme has a lower runtime overhead than the MPX
scheme as it avoids doing bound-checks
(Section~\ref{sec:spec-cpu}). However, it restricts the
segment size to $4$GB.

\noindent{\textbf{Multi-threading support.}}
Both our schemes support
multi-threading. All inserted runtime checks
(including those in Section~\ref{sec:cfi}) are thread-safe
because they check values of registers. However, we do need
additional support for thread-local storage (TLS). Typically, TLS is accessed via the segment register \ls{gs}: 
the base of TLS is obtained at a constant offset from \ls{gs}. 
The operating system takes care of setting \ls{gs} on a per-thread basis.
However, $\U$ and $\T$ operate in different trust domains, thus they cannot share the same TLS
buffer. 

We let $\T$ continue to use \ls{gs} for accessing its own TLS. 
\cc changes the compilation of $\U$ to access TLS in a different way. The
multiple (per-thread) stacks in $\U$ are all allocated inside the stack regions;
the public and private stacks for each thread are still at a constant offset to each
other. Each thread stack is, by default, of maximum size $1$MB and its start is
aligned to a $1$MB boundary (configurable at compile time). We keep the per-thread
TLS buffer at the beginning of the stack. $\U$ simply masks
the lower $20$-bits of \ls{rsp} to zeros to obtain the base of the stack and
access TLS. 

The segment-register scheme further requires switching of the \ls{gs} register
as control transfers between $\U$ and $\T$. We use appropriate wrappers to
achieve this switching, however $\T$ needs to reliably identify the current
thread-id when called from $\U$ (so that $\U$ cannot force two different threads
to use the same stack in $\T$). \cc achieves this by instrumenting an inlined-version
of the \ls{_chkstk}
routine\footnote{\url{https://msdn.microsoft.com/en-us/library/ms648426.aspx}} 
to make sure that \ls{rsp} does not escape its stack boundaries.



\section{Taint-aware CFI}
\label{sec:cfi}

We design a custom, taint-aware CFI scheme to ensure that an
attacker cannot alter the control flow of $\U$ to circumvent the
instrumented checks and leak sensitive data.

Typical low-level attacks that can hijack the control flow of a
program include overwriting the return address, or the targets of
function pointers and indirect jumps. Existing approaches
use a combination of \textit{shadow stacks} or \textit{stack
canaries} to prevent overwriting the return address, or use fine-grained
taint tracking to ensure that the value of a function pointer is not derived
from user (i.e. attacker-controlled) inputs \cite{Dang:2015:PCS:2714576.2714635,
Kuznetsov:2014:CI:2685048.2685061, pldi16}. While these techniques may prevent
certain attacks, our \emph{only} goal is ensuring confidentiality.
Thus, we designed a custom taint-aware CFI scheme. 

Our CFI scheme ensures that for each indirect transfer of control: 
(a) the target address is \textit{some} valid jump location, i.e., the target of
an indirect call is some valid procedure entry, and the target of a return is
some valid return site, and (b) the register taints expected at the target
address match the current register taints (e.g., when the \ls{rax}
register holds a private value then a \ls{ret} can only go to a
site that expects a \ls{private} return value). These suffice for our
goal of data confidentiality. Our scheme does not need to
ensure, for instance, that a return matches the previous call.

\noindent{\textbf{CFI for function calls and returns.}} We use a
\emph{magic-sequence} based scheme to achieve this CFI. We follow the 
x64 calling convention for Windows that has $4$ argument registers and
one return register. Our scheme picks two bit sequences $\Mcall$ and
$\Mret$ of length $59$ bits each that appear nowhere else in $\U$'s
binary. Each procedure in the binary is preceded with a string that
consists of $\Mcall$ followed by a $5$-bit sequence encoding the
expected taints of the $4$ argument registers and the return register,
as per the function signature. Similarly, each valid return site in
the binary is preceded by $\Mret$ followed by a $1$-bit encoding of
the taint of the return value register, again according to the
callee's signature. To keep the length of the sequences uniform at
$64$ bits, the return site taint is padded with four zero bits.
The $64$-bit instrumented sequences are collectively referred to as
magic sequences.

Callee-save registers are also live at function entry and
exit and their taints cannot be determined statically by the compiler.
\cc forces their taint to be public by making the caller save and clear all
the private-tainted callee-saved registers before making a \ls{call}. All dead
registers (e.g. unused argument registers and caller-saved
registers at the beginning of a function) are conservatively marked
\ls{private} to avoid accidental leaks.
We note that our scheme can be extended easily to
support other calling conventions.

Consider the following $\U$:
\begin{lstlisting}[frame=None,numbers=none]
private int add (private int x) { return x + 1; }
private int incr (private int *p, private int x) {
  int y = add (x); *p = y; return *p; }
\end{lstlisting}
The compiled code for these functions is instrumented
with magic sequences as follows. The $5$ taint bits for the \ls{add}
procedure are
\ls{11111} as its argument \ls{x} is \ls{private}, unused argument
registers are conservatively treated as \ls{private}, and its return
type is also \ls{private}. On the other hand, the taint bits for
\ls{incr} are \ls{01111} because its first argument is a \ls{public}
pointer (note that the \ls{private} annotation on the argument is on
the \ls{int}, not the pointer) , second argument is \ls{private},
unused argument registers are \ls{private}, and the return value is
also \ls{private}. For the return site in \ls{incr} after the call to
\ls{add}, the taint bits are \ls{00001} to indicate the private return
value register (with 4 bits of padding). The sample instrumentation is
as shown below:
\begin{lstlisting}[language={[x86masm]Assembler},frame=None,numbers=none]
#M_call#11111#
add: 
  ...  ;assembly code for add
#M_call#01111#
incr:
  ... ;assembly code of incr
  call add
  #M_ret#00001# ;private-tainted ret with padded 0s
  ... ;assembly code for rest of incr
\end{lstlisting}
Our CFI scheme adds runtime checks using these sequences as
follows. Each \ls{ret} is replaced with instructions to:
(a) fetch the return address, (b) confirm that the
target location has $\Mret$ followed by the taint-bit of the return
register, and if so, (c) jump to the target location. For our example,
the \ls{ret} inside \ls{add} is replaced as follows:
\begin{lstlisting}[language={[x86masm]Assembler},frame=None,numbers=none]
#M_call#11111#
add:
  ...
  r1 = pop ;fetch return address
  r2 = #M_ret_inverted#11110# ;bitwise negation 
  r2 = not r2                 ;of M_ret
  cmp [r1], r2                
  jne fail                    
  r1 = add r1, 8 ;skip magic sequence
  jmp r1 ;return
fail: call __debugbreak 
\end{lstlisting}
\begin{changebar}
We use the bitwise negation of $\Mret$ in the code to maintain the
invariant that the magic sequence does not appear in the binary at any
place other than valid return sites. There is no requirement that the
negated sequence not appear elsewhere in the binary.
\end{changebar}

For direct calls, \cc statically verifies that the register taints match between the call site
and the call target. At indirect calls, the instrumentation is
similar to that of a \ls{ret}: check that the target location contains $\Mcall$
followed by taint bits that match the register taints at the call site. 

\noindent{\textbf{Indirect jumps.}} \cc does not generate indirect
(non-call) jumps in $\U$.  Indirect jumps are mostly required for
jump-table optimizations, which we currently disable. We can
conceptually support them as long as the jump tables are statically
known and placed in read-only memory.


The insertion of magic sequences increases code size but it makes the
CFI-checking more lightweight than the shadow stack schemes. The unique
sequences $\Mcall$ and $\Mret$ are created post linking when the
binaries are available (Section~\ref{sec:ToolChain}).

\section{Implementation} \label{sec:compiler}

\newcommand{\code}[1]{\ls{#1}}

\begin{changebar}
We implemented \cc as part of the LLVM
framework~\cite{Lattner:2004:LCF:977395.977673}, targeting Windows and
Linux x64 platforms. It is possible to implement all of \cc's
instrumentation using simple branching instructions available on all
platforms, but we rely on x64's MPX support or x64's segment registers
to optimize runtime performance. We leave the optimizations on other
architectures as future work.
\end{changebar}

\subsection{\cc}

\noindent{\textbf{Compiler front-end.}} We introduce a new type qualifier,
\code{private}, in the language that the programmers can use to annotate
sensitive data. For example, a private integer-typed
variable can be declared as \code{private int x}, and a (public) pointer
pointing to a private integer as \code{private int *p}. The \code{struct}
fields inherit their \emph{outermost} annotation from the
corresponding \code{struct}-typed variable. For example, consider a
declaration \code{struct st \{ private int *p; \}}, and a
variable \code{x} of type \code{struct st}. Then \code{x.p} inherits its qualifier
from \code{x}: if \code{x} is declared as \code{private st x;}, then
\code{x.p} is a private pointer pointing to a private integer. 
\begin{changebar}
We follow the same convention with \code{union}s as well: all fields of a union 
inherit their outermost annotation from the \code{union}-typed variable.  

This convention ensures that despite the memory partitioning into public
and private regions, each object is laid out contiguously in memory in
only one of the regions. It does, however, carry the limitation that one cannot have 
structures  or unions whose fields have mixed outermost annotations, e.g., a 
\code{struct} with a \code{public int} field as well as a \code{private int} field, or a \code{union} 
over two structures, one with a \code{public int} field and another with a \code{private int} field. 
In all such cases, the programmer must restructure their code, often by simply introducing 
one level of indirection in the field (because the type constraints only apply to the outermost level).
\end{changebar}


We modified the Clang~\cite{Clang:URL} frontend to parse the
\private type qualifier and generate the LLVM 
Intermediate Representation (IR) instrumented with this additional
metadata.  Once the IR is generated, \cc runs standard LLVM IR
optimizations that are part of the LLVM toolchain. Most of the
optimizations work as-is and don't require any change. Optimizations
that change the metadata (e.g. \code{remove-dead-args} changes the function
signatures), need to be modified. While we found that it is not much
effort to modify an optimization, we chose to modify only the most
important ones in order to bound our effort. We disable the remaining
optimizations in our prototype.

\noindent{\textbf{LLVM IR and type inference.}}
After all the optimizations are run, our compiler runs a \emph{type
qualifier inference}~\cite{foster:pldi99} pass over the IR. This
inference pass propagates the type qualifier annotations to local
variables, and outputs an IR where all the intermediate variables are
optionally annotated with \private qualifiers. The inference is
implemented using a standard algorithm based on generating subtyping
constraints on dataflows, which are then solved using an SMT
solver~\cite{DeMoura:2008:ZES:1792734.1792766}. If the constraints are
unsatisfiable, an error is reported to the user. We refer the reader
to~\cite{foster:pldi99} for details of the algorithm.

After type qualifier inference, \cc knows the taint of each memory
operand for
\code{load} and \code{store} instructions. With a simple dataflow analysis
\cite{Aho:1986:CPT:6448}, the compiler statically determines the taint of each
register at each instruction. 


\noindent{\textbf{Register spilling and code generation.}} We made the
register allocator taint-aware: when a register is to be spilled on
the stack, the compiler appropriately chooses the private or the
public stack depending on the taint of the register. Once the LLVM IR
is lowered to machine IR, \cc emits the assembly code inserting all
the checks for memory bounds and taint-aware CFI.

\noindent{\textbf{MPX Optimizations.}}
\cc optimizes the bounds-checking in the MPX scheme. 
MPX instruction operands are identical to x64 memory operands, 
therefore one can check bounds of a complex operand using a single instruction. 
However, we found that bounds-checking a register is faster than
bounds-checking a memory operand (perhaps because using a memory operand requires an implicit {\tt lea}). 
\begin{changebar}
Consequently, \cc uses instructions on a register (as opposed to a
complex memory operand) as much as possible for bounds checking. 
\end{changebar}
It  reserves $1$MB of space around the public and private regions as guard regions
and eliminates the \textit{displacement} from the memory operand of a check 
if its absolute
value is smaller than $2^{20}$. 
Usually the displacement value is a small constant (for accessing structure
fields or doing stack accesses) and this optimization applies
to a large degree. Further, by enabling the \ls{_chkstk} enforcement
for the MPX scheme also (Section~\ref{sec:TaintEnforcement}), \cc eliminates
checks on stack accesses altogether because the \ls{rsp} value is
bound to be within the public region
(and \ls{rsp+OFFSET} is bound to be within the private region).

\cc further coalesces MPX checks within a basic block. Before adding a check,
it confirms if the same check was already added previously in the same
block, and there are no intervening \ls{call} instructions or 
subsequent modifications to the base or index registers.

\noindent{\textbf{Implicit flows.}}  By default, \cc tracks explicit
flows. It additionally produces a warning when the program branches on
private data, indicating the presence of a possible implicit
flow. Such a branch is easy to detect: it is a conditional jump on a
\private-tainted register.  Thus, if no such warning is produced, then
the application indeed lacks both explicit and implicit
flows. Additionally, we also allow the compiler to be used in an
all-\private scenario where all data manipulated by $\U$ is tainted
\private.  In such a case, the job of the compiler is easy: it only
needs to limit memory accesses in $\U$ to its own region of
memory. Implicit flows are not possible in this mode. None of our
applications (Section~\ref{sec:evaluation}) have implicit flows.

\subsection{\Verifier}
\label{sec:verifier}

$\Verifier$ checks that a binary produced by \cc has the required
instrumentation in place to guarantee that there are no (explicit)
private data leaks. The design goal of $\Verifier$ is to guard against
bugs in \cc; it is not meant for general-purpose verification of arbitrary
binaries.  $\Verifier$ actually helped us catch bugs in \cc during its
development.

$\Verifier$ is only 1500 LOC in addition to an off-the-shelf
disassembler that it uses for CFG construction. (Our current implementation
uses the LLVM disassembler.) This is three orders of magnitude smaller than
{\cc}'s $5$MLOC. Moreover, \Verifier is much simpler
than \cc; \Verifier does not include register allocation,
optimizations, etc.\ and uses a simple dataflow analysis to check all
the flows. Consequently, it provides a higher degree of assurance for
the security of our scheme.

\noindent{\textbf{Disassembly.}}
$\Verifier$ requires the unique prefixes of  magic sequences (Section~\ref{sec:cfi}) as input
and uses them  to identify procedure entries in the binary. It starts
disassembling the procedures and constructs their control-flow
graphs (CFG). $\Verifier$ assumes that the binary satisfies CFI, which makes it possible to
reliably identify all instructions in a procedure.
If the disassembly fails, the binary is rejected. Otherwise,
$\Verifier$ checks its assumptions: that the magic sequences were indeed unique in the procedures
identified and that they have enough CFI checks.

\noindent{\textbf{Data flow analysis and checks.}}
Next, $\Verifier$ performs a separate dataflow analysis on every
procedure's CFG to determine the taints of all the registers at each
instruction. It starts from the taint bits of the magic sequence
preceding the procedure.  It looks for MPX checks or the use of
segment registers to identify the taints of memory operands; if it
cannot find a check in the same basic block, the verification
fails. For each \ls{store} instruction, it checks that the taint of
the destination operand matches the taint of the source register. For
direct calls, it checks that the expected taints of the arguments, as
encoded in the magic sequence at the callee, matches the taints of the
argument registers at the callsite (this differs from \cc which uses
functions signatures). For indirect control transfers (indirect calls
and \ls{ret}), \Verifier confirms that there is a check for the magic
sequence at the target site and that its taint bits match the inferred
taints for registers. After a call instruction, \Verifier picks up
taint of the return register from the magic sequence (there should be
one), marks all caller-save registers as private, and
callee-save registers as public (following \cc's convention).

\noindent{\textbf{Additional checks.}}
$\Verifier$ additionally makes sure that a direct or a conditional
jump can only go a location in the same procedure.  $\Verifier$
rejects a binary that has an indirect jump, a system call, or if it
modifies a segment register.  $\Verifier$ also confirms correct usage
of \ls{_chkstk} to ensure that \ls{rsp} is kept within stack bounds.
For the segment-register scheme, $\Verifier$ additionally checks that each
memory operand uses only the lower $32$-bits of registers.

\noindent\textbf{Formal analysis.}
Although \Verifier is fairly simple, to improve our confidence in its
design, we built a formal model consisting of an abstract assembly
language with essential features like indirect calls, returns, as well
as \Verifier's checks. We prove formally that any program that
passes \Verifier's checks satisfies the standard information flow
security property of termination-insensitive
noninterference~\cite{Sabelfeld:2006:LIS:2312191.2314769}. In words,
this property states that data leaks are impossible. We defer the
details of the formalization to Appendix~\ref{appendix:A}.

\section{Toolchain}
\label{sec:ToolChain}

This section describes the overall flow to launch an application using our
toolchain. 

\noindent{\textbf{Compiling $\U$ using \cc.}} The only external
functions for $\U$'s code are those exported by $\T$. $\U$'s code is
compiled with an (auto-generated) stub file, that implements each of
these $\T$ functions as an indirect jump from a table \ls{externals},
located at a constant position in $\U$ (e.g.,
\ls{jmp (externals + offset)}$_i$
for the $i$-th function). The table \ls{externals} is initialized with
zeroes at this point, and \cc links all the $\U$ files to produce a
$\U$ dll.

The $\U$ dll is then post-processed to patch all the references to
globals, so that they correspond to the correct (private or public)
region. The globals themselves are relocated by the loader. The
post-processing pass also sets the $59$-bit prefix for the magic
sequences (used for CFI, Section~\ref{sec:cfi}). We find these
sequences by generating random bit sequences and checking for uniqueness; 
usually a small number of iterations are sufficient. 

\noindent{\textbf{Wrappers for $\T$ functions.}} For each of the functions in
$\T$'s interface exported to $\U$, we write a small wrapper that:
(a) performs the necessary checks for the arguments (e.g. the
\ls{send} wrapper would check that its argument buffer is
contained in $\U$'s public region), (b) copies arguments to $\T$'s
stack,
(c) switches \code{gs}, (d) switches \code{rsp} to $\T$'s
stack, and (e) calls the underlying $\T$ function
(e.g. \ls{send} in libc).  On return, it the wrapper switches \code{gs}
and \code{rsp} back and jumps to $\U$ in a similar manner to our CFI
return instrumentation.  Additionally, the wrappers include the magic
sequences similar to those in $\U$ so that the CFI checks in $\U$ do
not fail when calling $\T$.  These wrappers are compiled with the $\T$
dll, and the output dll exports the interface functions.

\noindent{\textbf{Loading the $\U$ and $\T$ dlls.}} When loading the $\U$ and
$\T$ dlls, the loader: $(1)$ populates the \ls{externals} 
table in $U$  with addresses of the wrapper functions in $\T$,
$(2)$ relocates the globals in $\U$ to their respective, private or
public regions, $(3)$ sets the MPX bound registers for the MPX scheme or the segment
registers for the segment-register scheme, and $(4)$ 
initializes the heaps and stacks in all the regions, marks them
non-executable, and jumps to the \ls{main} routine.

\noindent{\textbf{Memory allocator}}. \cc uses a customized memory 
allocator to enclose the private and public allocations in their 
respective sections. We modified dlmalloc~\cite{dlmalloc} to achieve this.

\section{Evaluation}
\label{sec:evaluation}
The goal of our evaluation is three-fold: (a) Quantify the performance
overheads of \cc's instrumentation, both for enforcing bounds and for
enforcing CFI; (b) Demonstrate that \cc scales to large, existing
applications and quantify changes to existing applications to make
them \cc-compatible; (c) Check that our scheme actually stops
confidentiality exploits in applications.

\subsection{CPU benchmarks}
\label{sec:spec-cpu}

We measured the overheads of \cc's instrumentation on the standard
SPEC CPU 2006 benchmarks~\cite{speccpu}. We treat the code of the
benchmarks as untrusted (in $\U$) and compile it with \cc. We use the
system's native libc, which is treated as trusted (in $\T$). These
benchmarks use no private data, so we added no annotations to the
benchmarks, which makes all data \ls{public} by default. Nonetheless,
the code emitted by \cc ensures that all memory accesses are actually
in the public region, it enforces CFI, and switches stacks when
calling $\T$ functions, so this experiment accurately captures \cc's
overheads.
We ran the benchmarks in the following configurations.
\begin{itemize}
\item[-] {\cBase}: Benchmarks compiled with vanilla LLVM,
  with O2 optimizations. This is the baseline for
  evaluation.\footnote{O2 is the standard optimization level
  for performance evaluation. Higher levels include
  ``optimizations'' that don't always speed up the program.}
\item[-] {\cBaseOA}:
Benchmarks compiled with \emph{vanilla} LLVM but running
with our custom allocator.
\item[-] {\cCC}: Compiled with {\cc}, but without any
  runtime instrumentation. However, all optimizations unsupported by
  {\cc} are disabled, the memories of $\T$ and $\U$ are separated and,
  hence, stacks are switched in calling $\T$ functions from $\U$.
\item[-] {\cCCcfi}: Like {\cCC} but additionally with CFI instrumentation,
  but no memory bounds enforcement.
\item[-] {\cCCmpx}: Full \cc, memory-bounds checks use MPX.
\item[-] {\cCCseg}: Full \cc, memory-bounds checks use segmentation.
\end{itemize}

Briefly, the difference between {\cCCcfi} and {\cCC} is the cost of
our CFI instrumentation. The difference between {\cCCmpx}
(resp. {\cCCseg}) and {\cCCcfi} is the cost of enforcing bounds using
MPX (resp. segment registers).

\begin{changebar}
We ran all of the C benchmarks of SPEC CPU 2006, except perlBench, which 
uses \ls{fork} that we currently do not support. 
\end{changebar}
All benchmarks were run on a Microsoft Surface Pro-4 Windows 10
machine with an Intel Core i7-6650U 2.20 GHz 64-bit processor with 2
cores (4 logical cores) and 8 GB RAM.

\begin{figure}
\includegraphics[width=1\linewidth]{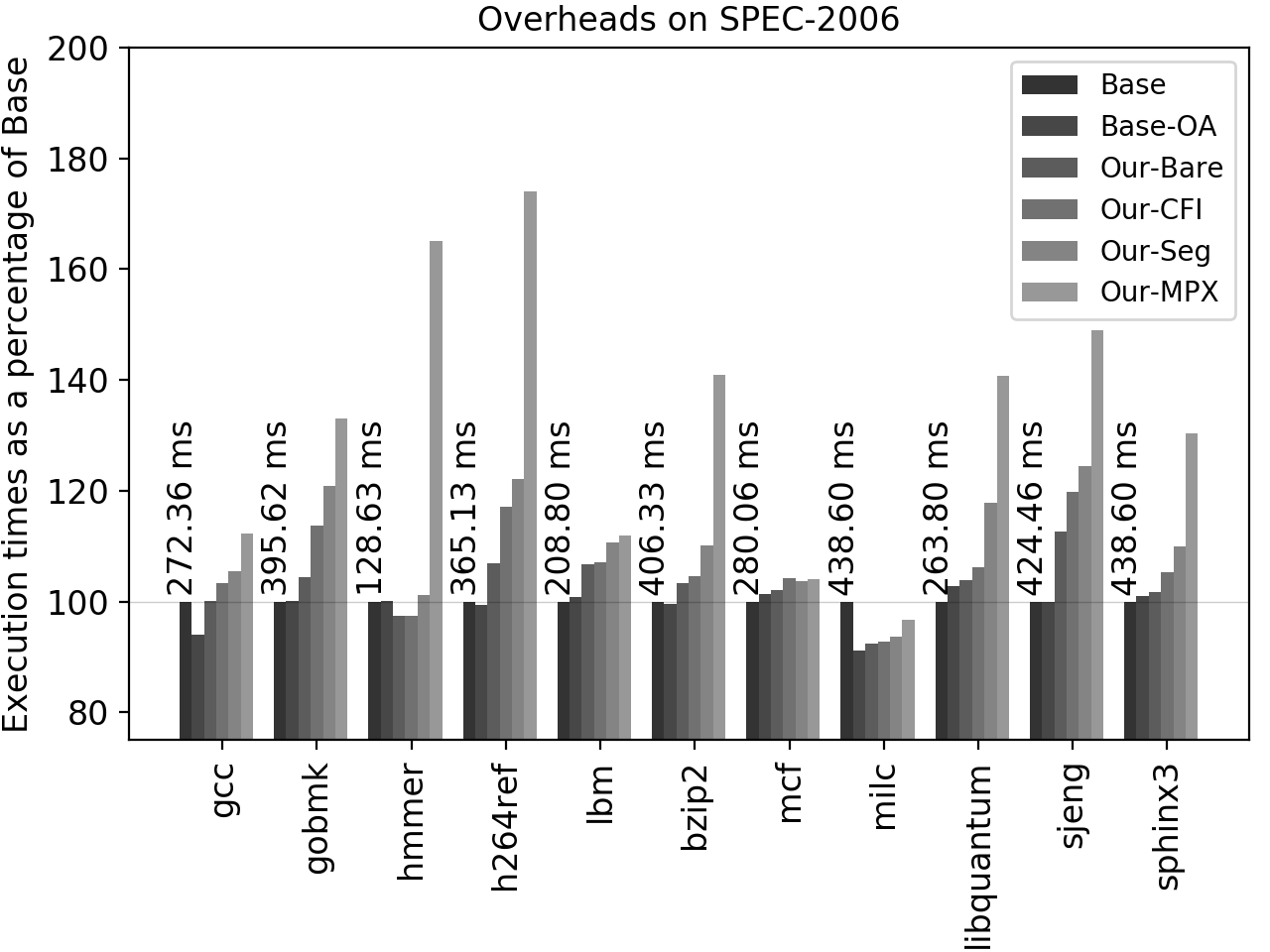}
\caption{\changed{Execution time as a percentage of {\cBase} for SPEC CPU 2006 benchmarks. Numbers above the {\cBase} bars are absolute execution times of the baseline in \ls{ms}. All bars are averages of 10 runs. Standard deviations are all below 3\%.}}
\label{table:spec-cpu}
\end{figure}

Figure~\ref{table:spec-cpu} shows
the results of our experiment.
The overhead of \cc using MPX (\cCCmpx) is up to 74.03\%, while that
of \cc using segmentation (\cCCseg) is up to 24.5\%.
As expected,
the overheads are almost consistently significantly lower when using
segmentation than when using MPX. Looking further, some of the
overhead (up to 10.2\%) comes from CFI enforcement ({\cCCcfi}$ -
${\cCC}). The average CFI overhead is $3.62\%$, competitive with best
known techniques \cite{Dang:2015:PCS:2714576.2714635}. Stack switching
and disabled optimizations (\cCC) account for the remaining
overhead. The overhead due to our custom memory allocator (\cBaseOA)
is negligible and, in many benchmarks, the custom allocator improves
performance.

We further comment on some seemingly odd results.  On mcf, the cost of
CFI alone ({\cCCcfi}, 4.17\%) seems to be higher than that of the full
MPX-based instrumentation ({\cCCmpx}, 4.02\%). We verified that this
is due to an outlier in the {\cCCcfi} experiment. On hmmer, the
overhead of {\cCC} is negative because the optimizations that {\cc}
disables actually slow it down.  Finally, on milc, the overhead of
{\cc} is negative because this benchmark benefits significantly from
the use of our custom memory allocator. Indeed, relative to
{\cBaseOA}, the remaining overheads follow expected trends.



\subsection{Web server: NGINX}

Next, we demonstrate that our method and {\cc} scale to large
applications. We cover three applications in this and the next two
sections. We first use {\cc} to protect logs in NGINX, the most
popular web server among high-traffic websites~\cite{nginx}. NGINX has
a logging module that logs time stamps, processing times, etc.\ for
each request along with request metadata such as the client address
and the request URI. An obvious confidentiality concern is that
sensitive content from files being served may leak into the logs due
to bugs. We use {\cc} to prevent such leaks.

We annotate NGINX's codebase to place OpenSSL in $\T$, and the rest of
NGINX, including all its request parsing, processing, serving, and
logging code in $\U$. The code in $\U$ is compiled with {\cc}
(total 124,001 LoC). Within $\U$, we mark everything as \ls{private},
except for the buffers in the logging module that are marked
as \ls{public}. To log request URIs, which are actually private, we
add a new \ls{encrypt_log} function to $\T$ that $\U$ invokes to
encrypt the request URI before adding it to the log. This function
encrypts using a key that is isolated in $\T$'s own region. The key is
pre-shared with the administrator who is authorized to read the
logs. The encrypted result of \ls{encrypt_log} is placed in
a \ls{public} buffer. The standard \ls{SSL_recv} function in $\T$
decrypts the incoming payload with the session key, and provides it to
$\U$ in a \ls{private} buffer.
\begin{lstlisting}[frame=non,numbers=none]
size_t SSL_recv(SSL *connection, 
       private void *buffer, size_t length);
\end{lstlisting}
\highlight{In total, we added or modified 160 LoC in NGINX (0.13\% of NGINX's
codebase) and added 138 LoC to $\T$.}


Our goal is to measure the overhead of {\cc} on the maximum sustained
throughput of NGINX. We run our experiments in 6 configurations:
{\cBase}, {\cCCnolib}, {\cCConlylib}, {\cCConlycfi}, and
{\cCCsamestack}, {\cCCfull}. Of these, {\cBase}, {\cCConlylib},
{\cCConlycfi}, and {\cCCfull} are the same as those in
Section~\ref{sec:spec-cpu}. We use MPX for bounds checks ({\cCCfull}),
as opposed to segmentation, since we know from
Section~\ref{sec:spec-cpu} that MPX is worse for {\cc}. {\cCCnolib} is
like {\cCConlylib} (compiled with {\cc} without any instrumentation),
but also does not separate memories for $\T$ and $\U$. {\cCCsamestack}
includes all instrumentation, but does not separate stacks for private
and public data. Briefly, the difference between {\cCConlylib} and
{\cCCnolib} is the overhead of separating $\T$'s memory from $\U$'s
and switching stacks on every call to $\T$, while the difference
between {\cCCfull} and {\cCCsamestack} is the overhead of increased
cache pressure from having separate stacks for private and public
data.

We host NGINX version 1.13.12 on an Intel Core i7-6700 3.40GHz 64-bit
processor with 4 cores (8 logical cores), 32 GB RAM, and a 10 Gbps
Ethernet card, running Ubuntu v16.04.1 with Linux kernel version
4.13.0-38-generic. 
\begin{changebar}
Hyperthreading was disabled in order to reduce experimental noise.
\end{changebar}
NGINX is configured to
run a single worker process pinned to a core. We connect to the server
from two client machines using the {\tt wrk2} tool~\cite{wrk2},
simulating a total of 32 concurrent clients. Each client makes 10,000
sequential requests, randomly chosen from a corpus of 1,000 files of
the same size (we vary the file size across experiments). The files
are served from a RAM disk. This saturates the CPU core hosting NGINX
in all setups.

\begin{figure}
\includegraphics[width=1\linewidth]{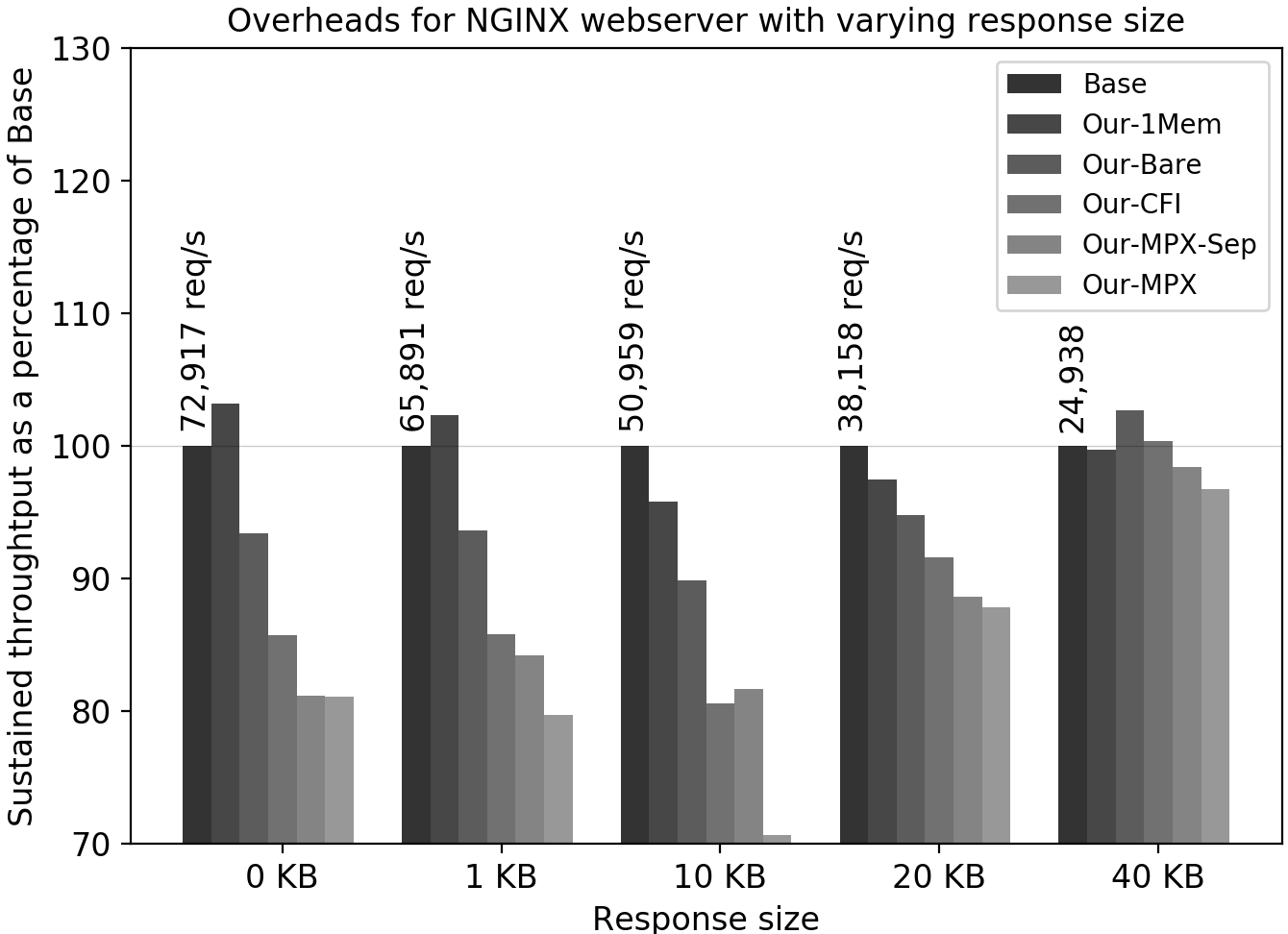}
\caption{\changed{Maximum sustained throughput as a percentage of {\cBase} for the NGINX web server with increasing response sizes. Numbers above the {\cBase} bars are absolute throughputs in \ls{req/s} for the baseline. All bars are averages of 10 runs. Standard deviations are all below 0.3\%, except in {\cBase} for response sizes 0 KB  and 1 KB, where they are below 2.2\%.}}
\label{table:nginx-res}
\end{figure}

Figure~\ref{table:nginx-res} shows the steady-state throughputs in the
six configurations for file sizes ranging from 0 to 40
KB.
For file sizes beyond 40 KB, the 10 Gbps
network card saturates in the base line before the CPU, and the excess
CPU cycles absorb our overheads.

Overall, {\cc}'s overhead on sustained throughput ranges from 3.25\% to
29.32\%. The overhead is not monotonic in file size or base line
throughput, indicating that there are compensating effects at
play. For large file sizes, the relative amount of time spent outside
$\U$, e.g., in the kernel in copying data, is substantial. Since code
outside $\U$ is not subject to our instrumentation, our relative
overhead falls for large file sizes ($>$10 KB here) and eventually
tends to zero. The initial increase in overhead up to file size 10 KB
comes mostly from the increased cache pressure due to the separation
of stacks for public and private data (the difference
{\cCCfull}$-${\cCCsamestack}). This is unsurprising: As the file size
increases, so does the cache footprint of $\U$. In contrast, the
overheads due to CFI (difference {\cCConlycfi}$-${\cCConlylib}) and
the separation of the memories of $\T$ and $\U$ (difference
{\cCConlylib}$-${\cCCnolib}) are relatively constant for small file
sizes.

Note that these overheads are moderate and we expect that they can be
reduced further by using segmentation in place of MPX for bounds
checks.



\subsection{OpenLDAP} \label{ssec:openldap}

Next, we apply {\cc} to OpenLDAP~\cite{openldap}, an implementation of
the Lightweight Directory Access Protocol (LDAP)~\cite{rfc4511}. LDAP
is a standard for organizing and accessing hierarchical
information. Here, we use {\cc} to protect root and user passwords
stored in OpenLDAP version 2.4.45. By default, the root password
(which authorizes access to OpenLDAP's entire store) and user
passwords are all stored unencrypted. We added new functions to
encrypt and decrypt these passwords, and modified OpenLDAP to use
these functions prior to storing and loading passwords,
respectively. The concern still is that OpenLDAP might leak in-memory
passwords without encrypting them. To prevent this, we treat all of
OpenLDAP as untrusted ($\U$), and protect it by compiling it with
{\cc}. The new cryptographic functions are in $\T$. Specifically,
decryption returns its output in a \ls{private} buffer, so {\cc}
prevents $\U$ from leaking it. The part we compile with {\cc}
is \ldaploc lines of C code, spread across \ldapfiles source
files. \highlight{Our modifications amount to \ldapadds new LoC for $\T$ and
$\ldapedits$ edited LoC in $\U$. Together, these constitute about
0.5\% of the original codebase.}


We configure OpenLDAP as a multi-threaded server (the default) with a
memory-mapped backing store (also the default), and simple
username/password authentication.  We use the same machine as for the
SPEC CPU benchmarks (Section~\ref{sec:spec-cpu}) to host an OpenLDAP
server configured to run 6 concurrent threads. The server is
pre-populated with 10,000 random directory entries. All memory-mapped
files are cached in memory before the experiment starts.

In our first experiment, 80 concurrent clients connected to the server
from another machine over a 100Mbps direct Ethernet link issue
concurrent requests for directory entries that do \emph{not}
exist. Across three trials, the server handles on average 26,254 and
22,908 requests per second in the baseline ({\cBase}) and {\cc} using
MPX ({\cCCmpx}). This corresponds to a throughput degradation of
12.74\%. The server CPU remains nearly saturated throughout the
experiment. The standard deviations are very small (1.7\% and 0.2\% in
{\cBase} and {\cCCmpx}, respectively).

Our second experiment is identical to the first, except that 60
concurrent clients issue small requests for entries that exist on the
server. Now, the baseline and {\cc} handle 29,698 and 26,895 queries
per second, respectively. This is a throughput degradation of
9.44\%. The standard deviations are small (less than 0.2\%).

The reason for the difference in overheads in these two experiments is
that OpenLDAP does less work in $\U$ looking for directory entries
that exist than it does looking for directory entries that don't
exist. Again, the overheads of {\cc}'s instrumentation are only
moderate and can be reduced further by using segmentation in place of
MPX.


\subsection{\cc with Intel SGX}
\label{sec:SGXapp}

Hardware features, such as the Intel Software Guard Extensions
(SGX)~\cite{sgx-explained}, allow \emph{isolating} sensitive code and
data into a separate \emph{enclave}, which cannot be read or written
directly from outside. Although this affords strong protection (even
against a malicious operating system),
bugs within the isolated code can still leak sensitive data out from the
enclave. This risk can be mitigated by using \cc to compile the code that
runs within the enclave.

\begin{changebar}
We experimented
with \textsc{Privado} \cite{DBLP:journals/corr/abs-1810-00602}, a
system that performs image classification by running a pre-trained
model on a user-provided image. \textsc{Privado} treats both the model
(i.e., its parameters) and the user input as sensitive information.
These appear unencrypted only inside an enclave. \textsc{Privado}
contains a port of Torch~\cite{th,thnn} made compatible with Intel SGX
SDK.  The image classifier is an eleven-layer neural network (NN) that
categorizes images into ten classes.

We treat both Torch and the NN as untrusted code $\U$ and compile them
using \cc.  We mark all data in the enclave \ls{private}, and map
$\U$'s private region to a block of memory within the enclave. $\U$
contains roughly 36K Loc.  The trusted code ($\T$) running inside the
enclave only consists of generic application-independent logic: the
enclave SDK, our memory allocator, stubs to switch between $\T$ and
$\U$, and a declassifier that communicates the result of image
classification (which is a private value) back to the host running
outside the enclave.  Outside the enclave, we deploy a web server that
takes encrypted images from clients and classifies them inside the
enclave.  The use of \cc crucially reduces trust on the
application-sensitive logic that is all contained in $\U$: \cc
enforces that the declassifier is the only way to communicate private
information outside the enclave.
\end{changebar}

\begin{figure}
\includegraphics[width=1\linewidth]{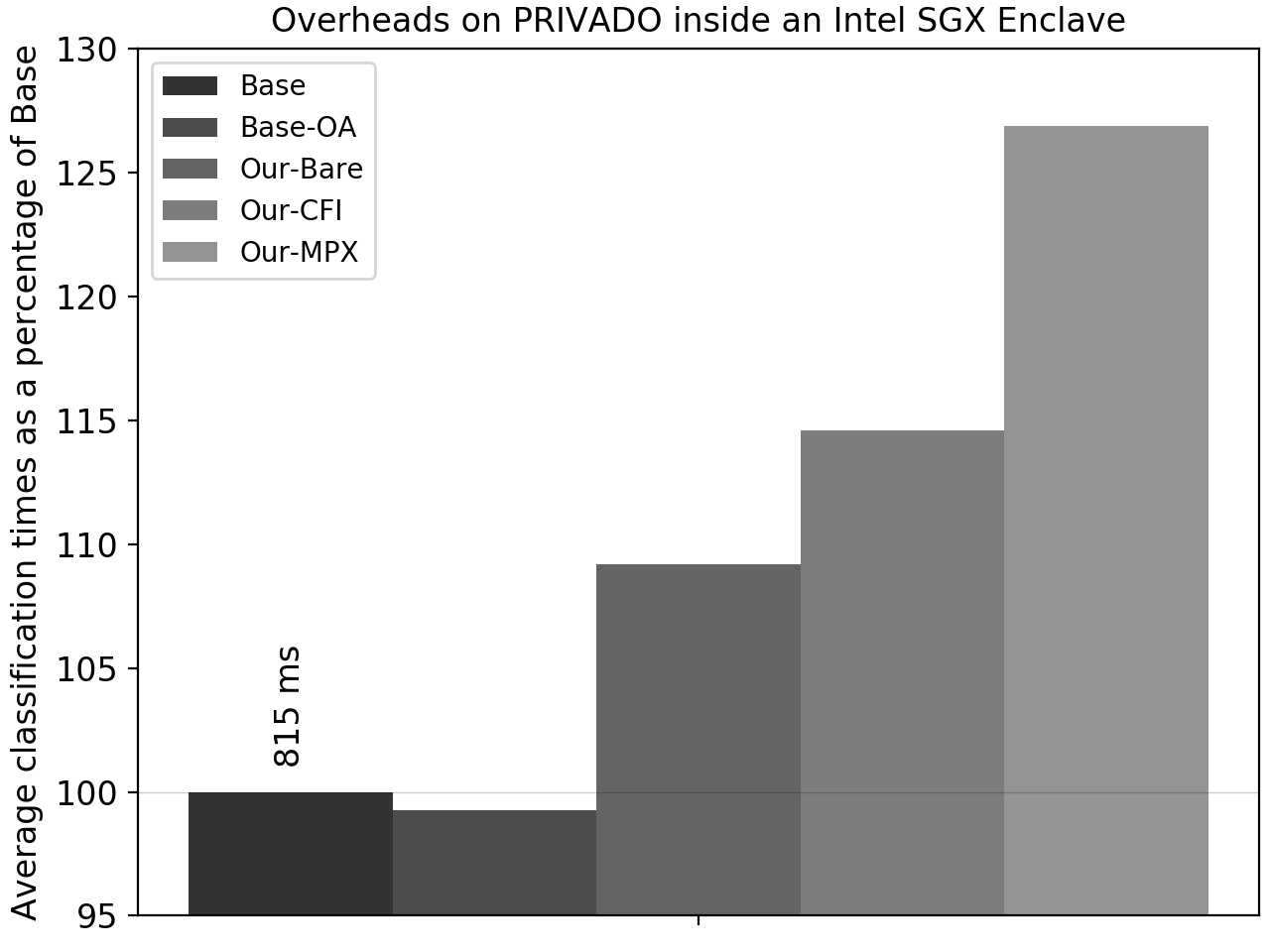}
\caption{\changed{Average classification time for \textsc{Privado} inside an Intel SGX Enclave as a percentage of {\cBase}. The number above the {\cBase} bar is the absolute execution time of the baseline in \ls{ms}. Every bar is the average of 10,000 trials. Standard deviations are all below 1\%.}}
\label{table:enclave}
\end{figure}

We ran this setup on an Intel Core i7-6700 3.40GHz 64-bit processor
with an Ubuntu 16.04 OS (Linux 4.15.0-34-generic kernel). 
We connect to our application from a client that issues 10,000
sequential requests to classify small (3 KB) files, and measure the
response time per image \emph{within the enclave} (thus excluding
latencies outside the enclave, which are common to the baseline and
our enforcement). We do this in 5 configurations of
Section~\ref{sec:spec-cpu}: \cBase, \cBaseOA, \cCC, {\cCCcfi}
and \cCCmpx. \begin{changebar}Figure~\ref{table:enclave} shows the average response
time for classifying an image in each of these five
configurations.\end{changebar}
For {\cBase}, we use the optimization level
O2, while the remaining configurations use O2 except
for two source files (out of 13) 
\begin{changebar}
where a bug in \cc (an optimization pass of O2 crashes)
\end{changebar}
forces us to use O0. This means that
the overheads reported in Figure~\ref{table:enclave} are higher than
actual \changed{and, hence, conservative}.

The overhead of {\cCCmpx} is 26.87\%. This is much lower than many of
the latency experiments of Section~\ref{sec:spec-cpu}. This is
because, in the classifier, a significant amount of time (almost 70\%)
is spent in a tight loop, which contains mostly only floating point
instructions and our instrumentation's MPX bound-check
instructions. These two classes of instructions can execute in
parallel on the CPU, so the overhead of our instrumentation is masked
within the loop.

\subsection{Data integrity and scaling with parallelism}
\label{sec:sys-libs}

The goal of our next experiment is two-fold: to verify that {\cc}'s
instrumentation scales well with thread parallelism, and to test that
{\cc} can be used to protect data \emph{integrity}, not just
confidentiality. We implemented a simple multi-threaded userspace
library that offers standard file read and write functionality, but
additionally provides data integrity by maintaining a Merkle hash tree
of file system contents. A security concern is that a bug in the
application or the library may clobber the hash tree to nullify
integrity guarantees. To prevent this, we compile both the library and
its clients using \cc (i.e. as part of $\U$). All data within the
client and the library is marked \ls{private}. The only exception is
the hash tree, which is marked \ls{public}. As usual, {\cc} prevents
the \ls{private} data from being written to \ls{public} data
structures accidentally, thus providing integrity for the hash
tree. To write hashes to the tree intentionally, we place the hashing
function in $\T$, allowing it to ``declassify'' data hashes
as \ls{public}.

\begin{figure}
\includegraphics[width=1\linewidth]{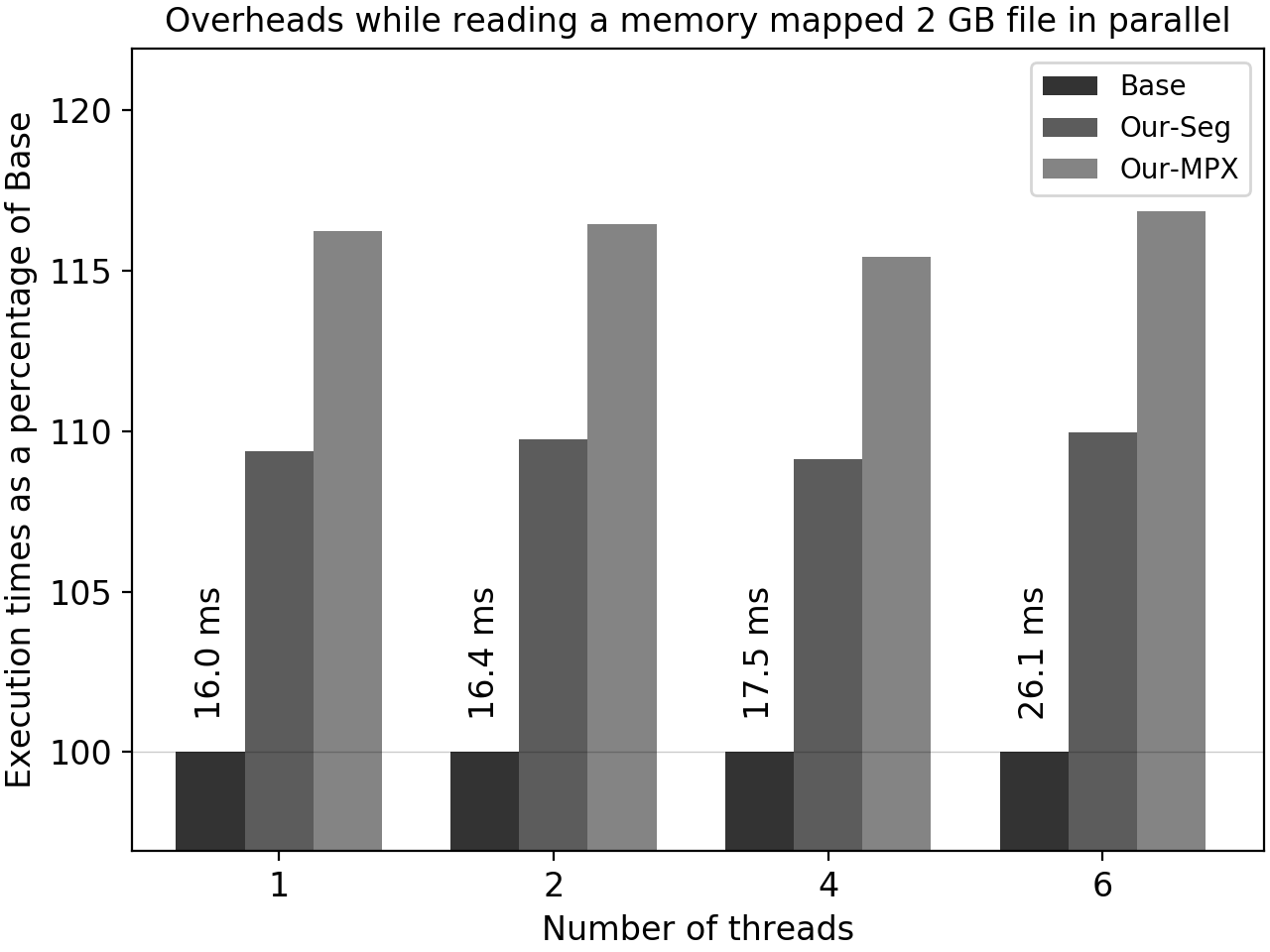}
\caption{\changed{Total execution time as a percentage of {\cBase} for reading a memory-mapped 2 GB file in parallel, as a function of the number of reading threads. Numbers above the {\cBase} bars are absolute execution times of the baseline in \ls{ms}. Each bar is an average of 5 runs. Standard deviations are all below 3\%.}}
\label{tab:libresults}
\end{figure}

We experiment with this library on a Windows 10 machine with an Intel
i7-6700 CPU (4 cores, 8 hyperthreaded cores) and 32 GB RAM.  Our
client program creates between 1 and 6 parallel threads, all of which
read a 2 GB file concurrently. The file is memory-mapped within the
library and cached previously. This gives us a CPU-bound workload. We
measure the total time taken to perform the reads in three
configurations: {\cBase}, {\cCCseg} and
{\cCCmpx}. \begin{changebar}Figure~\ref{tab:libresults} shows the total runtime as a function of the number of threads and the configuration.\end{changebar} Until the number of threads exceeds the number of cores (4), the
\changed{\emph{absolute} execution time (written above the {\cBase} bars)} and relative overhead of both the MPX and segmentation schemes
remains nearly constant, establishing linear scaling with the number
of threads. The actual overhead of {\cCCseg} is below 10\% and that of
{\cCCmpx} is below 17\% in all configurations.

\subsection{Vulnerability-injection experiments}

\begin{changebar}
To test that \cc stops data extraction vulnerabilities from being
exploited, we hand-crafted three vulnerabilities. In all cases,
compiling the vulnerable applications with \cc (after adding suitable
\ls{private} annotations) prevented the vulnerabilities from being
exploited.

First, we introduced a buffer-bounds vulnerability in the popular
Mongoose web server~\cite{mongoose} in the code path for serving an
unencrypted file over http. The vulnerability transmits any amount of
stale data from the stack, unencrypted. We wrote a client that
exploits the vulnerability by first requesting a private file, which
causes some contents of the private file to be written to the stack in
clear text, and then requesting a public file with the exploit. This
leaks stale private data from the first request. Using \cc on Mongoose
with proper annotations stops the exploit since \cc separates the
private and public stacks. The contents of the private file are
written to the private stack but the exploit reads from the public
stack.

Second, we modified Minizip~\cite{minizip}, a file compression tool,
to explicitly leak the file encryption password to a log file. \cc's
type inference detects this leak once we annotate the password
as \ls{private}. To make it harder for \cc, we added several pointer
type casts on the password, which make it impossible to detect the
leak statically. But, then, the dynamic checks inserted by \cc prevent
the leak.

Third, we wrote a simple function with a format string vulnerability
in its use of \ls{printf}. \ls{printf} is a vararg function, whose
first argument, the format string, determines how many subsequent
arguments the function tries to print. If the format string has more
directives than the number of arguments, potentially due to adversary
provided input,
\ls{printf} ends up reading other data from either the
argument registers or the stack. If any of this data is private, it
results in a data leak.
\cc prevents this vulnerability from being
exploited if we include \ls{printf}'s code in $\U$: since \ls{printf}
tries to read all arguments into buffers marked public, the bounds
enforcement of \cc prevents it from reading any private data.
\end{changebar}

\section{Discussion and Future Work}
\label{sec:limitations}

Currently, our scheme allows classifying data into two-levels---public
and private. It cannot be used for finer classification, e.g., to
separate the private data of Alice, the private data of Bob and public
data at the same time. We
also do not support label polymorphism at present, although that can
potentially be implemented using C++-like templates.


In our scheme, $\T$ is trusted and therefore must be implemented with
care. \cc guarantees that $\U$ cannot access $\T$'s memory or
jump to arbitrary points in $\T$,
so the only remaining attack surface for $\T$ is the API that it
exposes to $\U$. $\T$ must ensure that stringing together a sequence
of these API calls cannot cause leaks. We recommend the following
(standard) strategies. First, $\T$ should be kept small, mostly
containing application-independent functionality, e.g.,
communication interfaces, cryptographic routines, and optionally a
small number of libc routines (mainly for performance reasons), moving
the rest of the code to $\U$. Such a $\T$ can be re-used across
applications and can be subject to careful
audit/verification. Further, declassification routines in $\T$ must
provide guarded access to $\U$. For instance, $\T$ should disallow an
arbitrary number of calls to a password checking routine to prevent
probing attacks.



We rely on the absence of the magic sequence in $\T$'s binary to
prevent $\U$ from jumping inside $\T$.  We ensure
this by selecting the magic string when the entire code of $\U$ and
$\T$ is available. While dynamic loading in $\U$ can simply be disallowed, any
dynamic loading in $\T$ must ensure that the loaded library does not
contain the magic sequence. Since the ($59$-bits) magic sequence is
generated at random, the chances of it appearing in the loaded library
is minimal. A stronger
defense is to instrument indirect control transfers in $\U$ to remain
inside $\U$'s own code.

\cc supports callbacks from $\T$ to $\U$
with the help of \textit{trusted} wrappers in $\U$ that return to a
fixed location in $\T$, where $\T$ can restore its stack and start
execution from where it left off (or fail if $\T$ never called into
$\U$).

\begin{changebar}
At present, \cc only supports the C language. Implementing support for C++ remains 
as interesting future work. It requires integrating the private type qualifier 
with the C++ type system and ensuring that the C++ runtime and object system respects 
the user-intended taint flow. 
\end{changebar}



\section{Related Work}
\label{sec:related}

Our work bears similarities to Sinha et al.~\cite{pldi16} who proposed
a design methodology for programming secure enclaves (e.g., those that
use Intel SGX instructions for memory isolation).  The code inside an
enclave is divided into $\U$ and $\L$. $\U$'s code is compiled via a
special instrumenting compiler \cite{slashguard} while $\L$'s code is
trusted and may be compiled using any compiler. This is similar in
principle to our $\U$-$\T$ division. However, there are several
differences.  First, even the goals are different: their scheme does
not track taints; it only ensures that all unencrypted I/O done by
$\U$ goes through $\L$, which encrypts all outgoing data
uniformly. Thus, the application cannot carry out plain-text
communication even on public data without losing the security
guarantee.  Second, their implementation does not support
multi-threading (it relies on page protection to isolate $\L$ from
$\U$). Third, they maintain a \textit{bitmap} of writeable memory
locations for enforcing CFI, resulting in time and memory
overheads. Our CFI is taint-aware and without these
overheads. Finally, their verifier does not scale to SPEC benchmarks,
whereas our verifier is faster and scales to all
binaries that we have tried.

In an effort parallel to ours, Carr \emph{et al.}~\cite{datashield}
present DataShield, whose goal, like \cc's, is information flow
control in low-level code.
However, there are several differences between DataShield and our
work. First and foremost, DataShield itself only prevents
\emph{non-control} data flow attacks in which data is leaked or
corrupted without relying on a control flow hijack. A separate CFI
solution is needed to prevent leaks of information in the face of
control flow hijacks. In contrast, our scheme incorporates a
customized CFI that provides only the minimum necessary for
information flow control. One of the key insights of our work is that
(standard) CFI is neither necessary nor sufficient to prevent
information flow violations due to control flow hijacks. Second,
DataShield places blind trust in its compiler. In contrast, in our
work, the verifier \Verifier eliminates the need to trust the compiler
\cc. Third, DataShield enforces memory safety at object-granularity on
sensitive objects. This allows DataShield to enforce \emph{integrity}
for data invariants, which is mostly outside the scope of our
work. However, as we show in Section~\ref{sec:sys-libs}, our work can
be used to prevent untrusted data from flowing into sensitive
locations, which is a different form of integrity.

Rocha \emph{et al.}~\cite{rocha} and Banerjee
\emph{et al.}~\cite{stack-based} use combination of hybrid and static
methods for information flow control, but in memory- and type-safe
languages like Java. Cimplifier by Rastogi \emph{et
  al.}~\cite{cimplifier} tracks information flow only at the level of
process binaries by using separate docker containers.

Region-based memory partitioning has been explored before in the
context of safe and efficient memory
management~\cite{Tofte:1997:RMM:249657.249661,
  Grossman:2002:RMM:512529.512563}, but not for information flow.  In
\cc, regions obviate the need for dynamic taint
tracking~\cite{os-taint, straighttaint, explicitsecrecy}.
TaintCheck~\cite{taintcheck} first proposed the idea of dynamic taint
tracking, and forms the basis of
Valgrind~\cite{valgrind}. DECAF~\cite{decaf} is a whole system binary
analysis framework including a taint-tracking mechanism.  However,
such dynamic taint trackers incur heavy performance overhead. For
example, DECAF has an overhead of 600\%. Similarly, TaintCheck can
impose a 37x performance overhead for CPU-bound applications. Suh
\emph{et al.}~\cite{Suh:2004:SPE:1024393.1024404} report less than
$1\%$ overheads for their dynamic taint-tracking scheme, but they rely
on custom hardware.


Static analyses~\cite{vasan, lesscode, typesan, asan} of source code
can prove security-relevant criteria such as safe downcasts in C++, or
the correct use of variadic arguments. When proofs cannot be
constructed, runtime checks are inserted to enforce relevant policy at
runtime. This is similar to our use of runtime checks, but the
purposes are different.



Memory-safety
techniques for C such as
CCured~\cite{DBLP:journals/toplas/NeculaCHMW05} and
SoftBound~\cite{DBLP:conf/pldi/NagarakatteZMZ09} do not provide
confidentiality in all cases and already have overheads higher than
those of \cc (see~\cite[Section 2.2]{datashield} for a summary of
the overheads). Techniques such as
control flow integrity (CFI)~\cite{abadi2005control} and code-pointer
integrity (CPI)~\cite{Kuznetsov:2014:CI:2685048.2685061} prevent
control flow hijacks but not all data leaks. While our new taint-aware
CFI is an integral component of our enforcement, our goal of
preventing data leaks goes beyond CFI and CPI. Our CFI mechanism is
similar to Abadi et al.~\cite{abadi2005control} and Zeng et
al.~\cite{zeng2011combining} in its use of magic sequences, but our
magic sequences are taint-aware.




\bibliographystyle{plain}
\bibliography{references}

\appendix

\section{Formal model of verifier}
\label{appendix:A}



Conceptually, \Verifier operates in two stages. First, it disassembles
a binary, constructs its control flow graph (CFG) and re-infers the
taints of registers at all program points. Second, it checks that the
taints at the beginning and end of every instruction are consistent
with the instruction's semantics, and that other dynamic checks
described in Section~\ref{sec:verifier} are correctly inserted. Here,
we formalize the second stage and show if a program passes those
checks, then the program is secure in a formal sense. We assume that
the disassembly and the reconstruction of the CFG from the first
stage, both of which use an existing tool, are correct. 

\begin{table}[]
\small
\begin{equation*}
\begin{split}
\stmt \bnfdef &
  \bnfc{ldr} \of{\reg{}, exp} \ \                 \bnfsep 
  \bnfc{str} \of{\reg{}, exp} \                   \bnfsep 
  \bcjump{exp}                                    \bnfsep \\ &
  \beifthenelse{exp}{\bcjump{exp}}{\bcjump{exp}}  \bnfsep 
  \bcret{}                                        \bnfsep \\&
  \bccall{f}{exp^*}{\scaleto{\{\untrusted|\trusted\}}{4pt}}      \bnfsep
  \bicall{exp}{exp^*}                             \bnfsep 
  \bnfc{assert} \of{exp^*}                                                            
\\\\  
\mathit{exp} \bnfdef &
  n \in \values                         \             \bnfsep
  \reg{} \in \regs                      \             \bnfsep 
  \unaryop{exp}                         \             \bnfsep
  \binaryop{exp}{exp}                                 \bnfsep 
  \&f
\end{split}
\end{equation*}
\vspace{0.5cm}
\caption{Command syntax}
\label{tbl:bil_model_syntax}
\end{table}

We model the disassembled CFG abstractly. For each function, the CFG
has one bit of metadata, which indicates whether the function is part
of the trusted code or the untrusted code (abstractly written
$\trusted$ and $\untrusted$, respectively). Additionally, there is a
64-bit magic sequence for each function, which encodes the taints of
the function's arguments and its return value (see
Section~\ref{sec:cfi}). For trusted functions (whose code \Verifier
does not analyze), this is all the CFG contains. For untrusted
functions (whose code \Verifier does analyze), there is an additional
block-graph describing the code of the function. The block-graph for
an untrusted function $\meth$ is a tuple $\cfg_{\meth} =
\tuple{\cntrls_{\meth}, \edges}$ of nodes of $\meth$ and edges $\edges
\subseteq \cntrls \times \cntrls$. The edges represent all direct
control transfers. A node $\tuple{\pc, C, \seccontext, \seccontext'}
\in \cntrls_{\meth}$ consists of a program counter $\pc$, a single
command (assembly instruction) $C$, and register taints $\seccontext$
and $\seccontext'$ before and after the execution of $C$. $\pc$ is a
number, like a line number, that can be used in an indirect jump to
this node. $\seccontext$ and $\seccontext'$ are maps from machine
register ids to $\{\hseclabel, \lseclabel\}$, representing high
(private) and low (public) data.

Commands $C$ are represented in an abstract syntax shown in
Table~\ref{tbl:bil_model_syntax}. The auxiliary syntax of
\emph{expressions} consists of constants, which here can only be
integers that may represent ordinary data or line numbers, standard
binary and unary operators (ranged over by $\binaryop{}{}$ and
$\unaryop{}$, respectively), and the special operation $\&f$ which
returns the $\pc$ of the starting instruction of function $f$. We
assume that operations used in expressions are total, i.e.,
expressions never get stuck.

Commands include register load $\bnfc{ldr} \of{\reg{}, exp}$ and store
$\bnfc{str} \of{\reg{}, exp}$ where the input $exp$ evaluates to a
memory location, unconditional jumps $\bcjump{exp}$, conditionals
$\bnfc{ifthenelse}$, direct function calls
$\bccall{f}{exp^*}{\scaleto{\{\untrusted|\trusted\}}{4pt}}$, indirect
function calls $\bnfc{icall}$, and return $\bcret$. The subscript for
$\bnfc{call}_{\scaleto{\{\untrusted|\trusted\}}{4pt}}$ denotes if the
command is used to invoke an untrusted functionality or a trusted
one. An additional command $\bnfc{assert}$ models checks inserted by
the compiler. The asserted expressions must be true, else the program
halts in a special state $\bot$. Figure~\ref{apx:symsem} presents the
operational semantic of the commands in
Table~\ref{tbl:bil_model_syntax}.

Every CFG $\cfg$ has a designated entry function, similar to main() in
C programs.

\noindent{\textbf{Dynamic semantics.}}
A program/CFG $\cfg$ can be evaluated in a context of a data memory
$\mem:\!\values\!\to\!\values$, a register state
$\regstate:\!\regs\!\to\!\values$ and a program counter $\pc$ which
points to a node in the program's CFG. The memory $\mem$ is actually a
union of two maps $\mem_{\lseclabel}$ and $\mem_{\hseclabel}$ over
disjoint domains, representing the low and the high regions of memory
of our scheme.
%
%
Given a fixed $\cfg$, a \textit{configuration} $\config$ of $\cfg$ is
a record $\tuple{\mem, \regstate,  [\stack_{\hseclabel}:\stack_{\lseclabel}], \pc}$ of the entire
state---the memory, register state, stack, and the program counter. We
make a distinction between high ($\stack_{\hseclabel}$) and low
($\stack_{\lseclabel}$) stacks. We write $\elem{\tuple{.}}{i}$ to
project a tuple to its $i$th component. Thus, if $\config =
\tuple{\mem, \regstate, [\stack_{\hseclabel}:\stack_{\lseclabel}], \pc}$,
then $\config.\pc = \pc$. We call a configuration $\config$ initial
(final) if $\config.\pc$ is the first (last) node of the entry
function of $\cfg$.


Next, we define the dynamic semantics of a program as a transition
relation $\config \to \config'$ (Figure~\ref{fig:opsemantics}).
We use $e\evaluates \mvalue$ to mean evaluation of an expression to its corresponding (concrete) value.
Moreover, we extend the  configuration $\config$ of $\cfg$ with an additional component $\tstate$ that represents the memory fragment available only to the trusted code and its protected against untrusted accesses, i.e.,
$\tuple{\tstate, \smcstate{},[\stack_{\hseclabel}: \stack_{\lseclabel}], \pc}$. Further, we introduce the record $\funstruct = \{f_i \mapsto \tuple{n_i,\cmagicn_i}\ |\ f_i \in \prog \}$ and extend  
the typing judgment with this record $\funstruct, \cfg
\vdash\seccontext \{pc\} \seccontext'$. $\funstruct$ keeps for each
function its starting node in the CFG and the magic sequence
associated with that
function. We only note that calls to
trusted functions, whose code is not modeled in the CFG, are
represented via an external relation $\config \hookrightarrow_{\meth}
\config'$ that models the entire execution of the trusted function
$\meth$ and its effects on the configuration. The special transition
$\config \to \bot$ means that $\config.\pc$ contains an
$\bnfc{assert}$ command, one of whose arguments evaluates to
false. $\bot$ does not transition any further. 
Furthermore, the transition $\config \to \confgiAdversary$ is used to model adversarial behavior, for example,
when the  target of $\bnfc{icall}$ is not in the CFG.
$\confgiAdversary$ resembles a configuration where memory locations are loaded with random values.


%

\noindent{\textbf{{Security Analysis.}}
We formalize the checks that \Verifier makes via the judgment $\cfg
\vdash\seccontext \{pc\} \seccontext'$, which means that the command
in the node labeled $\pc$ in the CFG $\cfg$ is consistent with the
beginning and ending taints $\seccontext$ and $\seccontext'$, and that
the checks from Section~\ref{sec:verifier} corresponding to this
command are satisfied. Rules for this judgment are shown in
Figure~\ref{apx:symsem}. The function $\mathit{pred}(\cfg, \pc)$ returns
the predecessors of the node labeled $\pc$ in $\cfg$. For $\seclabel
\in \{\lseclabel, \hseclabel\}$, $\bnfc{assert}(e \in
\preimg(\mem_{\seclabel}))$ represents the dynamic check that $e$
evaluates to an address in the domain of $\mem_{\seclabel}$, and the
auxiliary judgment $\seccontext \vdash e : \seclabel$ means that the
expression $e$ depends only on values with secrecy $\seclabel$ or
lower.

The rules are mostly self-explanatory. The first rule, which is for
the command $\bnfc{ldr}(\mathit{reg}, e)$, says that if $\mathit{reg}$
has taint $\seclabel_e$ after the command, then on all paths leading
to this command, there must be a check that whatever $e$ evaluates to
is actually pointing into $\mem_{\seclabel_e}$. The rule for
$\bnfc{str}$ is similar. In the rules for indirect branching, we
insist that the addresses of the branch targets have the taint
$\lseclabel$. Overall, our type system's design is inspired by the
flow-sensitive information flow type system of Hunt and
Sands~\cite{Hunt:2006:FST:1111037.1111045}. However, the runtime
checks are new to our system.

\newcommand{\cons}[1]{\underset{#1}{\mathit{Cons}}}

We formally define call- and return-sites magic sequences based on the CFG structure.
 Let $\cons{}$ be the bit-concatenation function, then:
(i) for the function entry node $\cntrl$, $\cmagicn = \cons{i = 0\dots3}(\cntrl.\seccontext(\reg{i}))$, and
(ii) for a given return address $adr$ and the node $\cntrl$ such that $\cntrl \in \pred{\cfg}{adr}$, $\rmagicn = \cntrl.\seccontext'(\reg{0})$.
We now turn to explain the meaning of rules in Figure~\ref{apx:symsem}.

\paragraph{$\bnfc{call}$} For the $\bnfc{call}$ statement we check that the expected taints
of the arguments, as encoded in the magic sequence at the callee, matches the taints of the argument registers at the callsite. It is also worth noting that at runtime all return addresses
are stored in the stack allocated for the low-security context.

\paragraph{$\bnfc{icall}$} For the indirect calls \textsc{ConfVerify} confirms that the function pointer is low and that there is a check for the magic
sequence at the target site and its taint bits match the inferred taints for registers. Note that since the check 
$\bnfc{assert}(\secle{\cmagicn}{\seclabel_{[1-4]}} \textbf{ and }  e_f \in \cfg \textbf{ and } ({e_f}\!\mapsto\!\cmagicn)\!\in\!\funstruct)$ is a runtime condition, the pointer ${e_f}$ will be evaluated to the corresponding function name at the execution
time and we can retrieve the magic string directly from $\funstruct$. The condition $e_f \in \cfg$ ensures that the target of the call statement is a valid node in the CFG of the program.

\paragraph{$\bnfc{ret}$} Similar to indirect function call, for the $\bnfc{ret}$ command we check that there is a check for the magic
sequence at the target site and that its taint bits match the inferred taints for registers. Additionally, \textsc{ConfVerify} confirms that 
the return address on the stack has a magic signature that has a taint bit of inferred return type $\seclabel$ for the function. In this rule we 
use the function $\mathit{top}$ with the standard meaning to manipulate the stack content. Again since 
$\bnfc{assert}(\forall \cntrl'\!\in \pred{\cfg}{\mathit{top}(\stack_{\lseclabel})}.\ \secle{\cntrl'.\seccontext'(\reg{0})}{\seclabel})$ is checked at runtime, we will have access to the stack.

\paragraph{$\bnfc{ifthenelse}$} For this rule we require that the security level of conditional expression is not $\hseclabel$.
Checks on the $\bnfc{goto}$ and $\bnfc{ifthenelse}$ guarantee that the program flow is secret independent.

We say that a CFG $\cfg$ is well-typed (passes \Verifier's checks),
written $\vdash \cfg$, when two conditions hold for all nodes $\cntrl$
in (untrusted functions in) $\cfg$: 1) The node (locally) satisfies
the type system. Formally, $\cfg \vdash \cntrl.\seccontext \;
\{\cntrl.\pc\}\; \cntrl.\seccontext'$, and 2) The ending taint of the
node is consistent with the beginning taints of all its
successors. Formally, for all nodes $\cntrl' \in \mathit{succ}(\cfg,
\cntrl.\pc)$, $\cntrl.\seccontext' \sqsubseteq \cntrl'.\seccontext$,
where $\mathit{succ}(\cfg, \pc)$ returns the successors of the node
labeled $\pc$ in $\cfg$. We emphasize again that only untrusted
functions are checked.

\noindent{\textbf{Security theorem.}}
The above checks are sufficient to prove our key theorem: If a program
passes \Verifier, then assuming that its trusted functions don't leak
private data, the whole program does not leak private data,
end-to-end. We formalize non-leakage of private data as the standard
information flow property called \emph{termination-insensitive
  noninterference}. Roughly, this property requires a notion of low
equivalence of program configurations (of the same CFG $\cfg$),
written $\config \leqv \config'$, which allows memories of $\config$
and $\config'$ to differ only in the private region.  A program is
noninterfering if it \emph{preserves} $\leqv$ for any two runs, except
when one program halts safely (e.g., on a failed
assertion). Intuitively, noninterference means that no information
from the private part of the initial memory can leak into the public
part of the final memory.

For our model, we define $\config \leqv \config'$ to hold if: (i)
$\config$ and $\config'$ point to the same command, i.e., $\config.\pc
= \config'.\pc$, (ii) the contents of their low-stacks are equal,
$\config.\stack_{\lseclabel} = \config'.\stack_{\lseclabel}$, (iii)
for all low memory addresses $m \in \mem_{\lseclabel}$,
$\config.\mem(m) = \config'.\mem(m)$, (iv) for all registers $r$ such
that $\cfg(\config.\pc).\seccontext(r) = \lseclabel$,
$\config.\regstate(r) = \config'.\regstate(r)$.


The assumption that trusted code does not leak private data is
formalized as follows.

\begin{assumption}
For all $\config_0$, $\config_1$, $\config'_0$ such that $\config_0
\leqv \config_1$, if $\config_0 \hookrightarrow_{\meth} \config'_0$
then $\exists \config'_1. \ \config_1 \hookrightarrow_{\meth}
\config'_1$ and $\config'_0 \leqv \config'_1$.
\end{assumption}

Under this assumption on the trusted code, we can show the
noninterference or security theorem. 
A necessary condition to show noninterference, however, is to ensure that no well-typed program can reach an ill-formed configuration.

\begin{lemma}\label{lem:noAdversarialConfig}
 Suppose $\vdash \cfg$, for all configurations $\config$ of $\cfg$  it holds that $\config \nrightarrow^* \confgiAdversary$.
\end{lemma}

Lemma~\ref{lem:noAdversarialConfig} rules out possible nondeterminism caused by adversarial behavior and allows to formalize the security theorem as follows.

\begin{theorem}[Termination-insensitive noninterference]
Suppose $\vdash \cfg$. Then, for all configurations $\config_0$,
$\config_0'$ and $\config_1$ of $\cfg$ such that $\config_0 \leqv
\config_1$ and $\config_0 \to^* \config'_0$, then either $\config_1
\to^* \bot$ or $\exists \config'_1.\ \config_1 \to^* \config'_1$ and
$\config'_0 \leqv \config'_1$.
\end{theorem}

When $\config_0$ and $\config_0'$ are initial and final
configurations, respectively, then $\config_1$ and $\config_1'$ must
also be initial and final configurations, so the theorem guarantees
freedom from data leaks, end-to-end (modulo assertion check failures).

\begin{figure}
\centering
{\footnotesize
\begin{math}
\prooftree
\begin{array}{l}
  C = \bnfc{ldr}(\reg{},e)\ \ 
  \smcstate{} \vdash e\evaluates \mvalue \ \
  \mvalue \in \left(\preimg(\mem_{\lseclabel})\!\cup\!\preimg(\mem_{\hseclabel})\right)
\end{array}
\justifies
\begin{array}{l}
 \tuple{\tstate, \smcstate{},[\stack], \pc} \to 
 \tuple{\tstate, \mem, \regstate[{\reg{}} \mapsto {\mem(\cntrl)}],[\stack], \pc+1}
\end{array}
\endprooftree 
\end{math}\\[10pt]
 
\begin{math}
\prooftree
\begin{array}{l}
  C = \bnfc{ldr}(\reg{},e)\ \ 
  \smcstate{} \vdash e\evaluates \mvalue\ \ 
  \mvalue \notin \left(\preimg(\mem_{\lseclabel})\!\cup\!\preimg(\mem_{\hseclabel})\right)
\end{array}
\justifies
\begin{array}{l}
 \tuple{\tstate, \smcstate{},[\stack], \pc} \to \confgiAdversary
\end{array}
\endprooftree 
\end{math}\\[10pt]

\begin{math}
\prooftree
\begin{array}{c}
  C = \bnfc{str}(\reg{},e)\ \   
  \smcstate{} \vdash e\evaluates {\mvalue}\ \ 
  \mvalue \in \left(\preimg(\mem_{\lseclabel})\!\cup\!\preimg(\mem_{\hseclabel})\right)
\end{array}
\justifies
\begin{array}{l}
 \tuple{\tstate, \smcstate{},[\stack], \pc} \to 
 \tuple{\tstate, \mem[{\cntrl} \mapsto {\regstate(\reg{})}], \regstate,[\stack], \pc+1}
\end{array}
\endprooftree 
\end{math}\\[10pt]

\begin{math}
\prooftree
\begin{array}{c}
  C = \bnfc{str}(\reg{},e)\ \  
  \smcstate{} \vdash e\evaluates {\mvalue}\ \ 
  \mvalue \notin \left(\preimg(\mem_{\lseclabel})\!\cup\!\preimg(\mem_{\hseclabel})\right)
\end{array}
\justifies
\begin{array}{l}
 \tuple{\tstate, \smcstate{},[\stack], \pc} \to \confgiAdversary
\end{array}
\endprooftree 
\end{math}\\[10pt]

\begin{math}
\prooftree
\begin{array}{c}
  C = \bccall{f_u}{e_1, \dots, e_4}{\untrusted}\ \ ({f_u} \mapsto \tuple{{\pc_f}, -})  \in \funstruct 
\end{array}
\justifies
\begin{array}{c}
 \tuple{\tstate, \smcstate{},[\stack],  \pc} \to 
 \tuple{\tstate, \smcstate{},[(\pc+1);\stack_{\lseclabel}: \stack_{\hseclabel}], {\pc}_f}
\end{array}
\endprooftree
\end{math}\\[10pt]

\begin{math}
\prooftree
\begin{array}{c}
  C = \bccall{f_{\tau}}{e_1, \dots, e_4}{\trusted}\ \ ({f_{\tau}} \mapsto \tuple{{\pc_f}, -})  \in \funstruct 
\end{array}
\justifies
\begin{array}{c}
 \tuple{\tstate, \smcstate{},[\stack],  \pc} \tto_{f_{\tau}}
 \tuple{\tstate', \mem', \regstate',[(\pc+1);\stack'_{\lseclabel}: \stack'_{\hseclabel}], \pc_f}
\end{array}
\endprooftree
\end{math}\\[10pt]

\begin{math}
\prooftree
\begin{array}{c}
  C = \bnfc{icall} {e_f}\of{e_1, \dots, e_4}\ \ \smcstate{} \vdash e_f\evaluates {\pc_f}\ \ \pc_f \in \cfg
\end{array}
\justifies
\begin{array}{c}
 \tuple{\tstate, \smcstate{},[\stack],  \pc} \to 
 \tuple{\tstate, \smcstate{},[(\pc+1);\stack_{\lseclabel}: \stack_{\hseclabel}], \pc_f}
\end{array}
\endprooftree
\end{math}\\[10pt]

\begin{math}
\prooftree
\begin{array}{c}
  C = \bnfc{icall} {e_f}\of{e_1, \dots, e_4}\ \ \smcstate{} \vdash e_f\evaluates {\pc_f}\ \ \pc_f \notin \cfg
\end{array}
\justifies
\begin{array}{c}
 \tuple{\tstate, \smcstate{},[\stack],  \pc} \to \confgiAdversary
\end{array}
\endprooftree
\end{math}\\[10pt]

\begin{math}
\prooftree
\begin{array}{l}
  C = \bcret\ \ \ adr \in \cfg
\end{array}
\justifies
\begin{array}{l}
 \tuple{\tstate, \smcstate{},[adr;\stack_{\lseclabel}\!:\!\stack_{\hseclabel}], \pc} \to 
 \tuple{\tstate, \smcstate{},[\stack_{\lseclabel}\!:\!\stack_{\hseclabel}], \mathit{adr}]}
\end{array}
\endprooftree 
\end{math}\\[10pt]

\begin{math}
\prooftree
\begin{array}{l}
  C = \bcret \ \ \ adr \notin \cfg
\end{array}
\justifies
\begin{array}{l}
 \tuple{\tstate, \smcstate{},[adr;\stack_{\lseclabel}\!:\!\stack_{\hseclabel}], \pc} \to \confgiAdversary
\end{array}
\endprooftree 
\end{math}\\[10pt]

\begin{math}
\prooftree
\begin{array}{c}
  C = \bcjump{e}\ \ \  \smcstate{} \vdash e \evaluates \pc
\end{array}
\justifies
\begin{array}{c}
 \tuple{\tstate, \smcstate{},[\stack], \pc} \to 
 \tuple{\tstate, \smcstate{},[\stack], \pc]}
\end{array}
\endprooftree 
\end{math}\\[10pt]

\begin{math}
\prooftree
\begin{array}{c}
  C = \beifthenelse{e}{\bcjump{e_1}}{\bcjump{e_2}}\ \ \smcstate{} \vdash e \evaluates T\ \   \smcstate{} \vdash e_1 \evaluates \mvalue_1
\end{array}
\justifies
\begin{array}{c}
 \tuple{\tstate, \smcstate,[\stack], \pc} \to 
 \tuple{\tstate, \smcstate{},[\stack], \mvalue_1}
\end{array}
\endprooftree 
\end{math}\\[10pt]

\begin{math}
\prooftree
\begin{array}{c}
  C = \beifthenelse{e}{\bcjump{e_1}}{\bcjump{e_2}}\ \ \smcstate{} \vdash e \evaluates F\ \ \smcstate{} \vdash e_2 \evaluates \mvalue_2
\end{array}
\justifies
\begin{array}{c}
 \tuple{\tstate, \smcstate{},[\stack], \pc} \to 
 \tuple{\tstate, \smcstate{},[\stack], \mvalue_2}
\end{array}
\endprooftree 
\end{math}\\[10pt]

\begin{math}
\prooftree
\begin{array}{c}
  C = \bnfc{assert}(e)\ \ \  \smcstate{} \vdash e \evaluates T
\end{array}
\justifies
\begin{array}{c}
 \tuple{\tstate, \smcstate{},[\stack], \pc} \to 
 \tuple{\tstate, \smcstate{},[\stack], \pc + 1}
\end{array}
\endprooftree 
\end{math}\\[10pt]

\begin{math}
\prooftree
\begin{array}{c}
  C = \bnfc{assert}(e)\ \ \  \smcstate{} \vdash e \evaluates F
\end{array}
\justifies
\begin{array}{c}
 \tuple{\tstate, \smcstate{},[\stack], \pc} \to \bot
\end{array}
\endprooftree 
\end{math}

}
\caption{Operational semantics rules.} 
\label{fig:opsemantics}
\end{figure}

\begin{figure}[ht]
\centering
{\footnotesize
\begin{tabular}{c}

\begin{math}
\prooftree
\begin{array}{c}
 C\!=\!\bnfc{ldr}(\reg{},e) \\  \forall\ \cntrl \!\in\!\pred{\cfg}{\pc}.\ \elem{\cntrl}{C}\!=\!\bnfc{assert}(e \in \preimg(\mem_{\seclabel_e})) 
\end{array}
\justifies
\begin{array}{c}
\funstruct,  \cfg \vdash \seccontext \{\pc\} \seccontext[reg \mapsto \seclabel_e]
\end{array}
\endprooftree  
\end{math}
\\[20pt]

\begin{math}
\prooftree
\begin{array}{c}
 C\!=\!\bnfc{str}(\reg{},e)\ \ \ \  \seccontext \vdash \reg{}: \seclabel_r \ \ \ \ \seclabel_r  \sqsubseteq \seclabel_{e}\\[5pt]
 \forall \cntrl\!\in\!\pred{\cfg}{\pc}.\ \elem{\cntrl}{C}\!=\!\bnfc{assert}(e\!\in\!\preimg(\mem_{\seclabel_{e}}))
\end{array}
\justifies
\begin{array}{l}
\funstruct,  \cfg \vdash \seccontext \{\pc\} \seccontext
\end{array}
\endprooftree  
\end{math}
\\[20pt]

\begin{math}
\prooftree
\begin{array}{c}
C = \bccall{f}{e_1, \dots, e_4}{\scaleto{\{\untrusted|\trusted\}}{4pt}}\ \ ({f}\!\mapsto\!\tuple{-,\cmagicn})  \in \funstruct\\[5pt]
\seccontext \vdash e_i : \seclabel_i\ \ \ \text{ for }\ \ i = 1 \dots 4 \ \ \ \ \ \ \ \seclabel_{[1-4]} \sqsubseteq \cmagicn
\end{array}
\justifies
\begin{array}{l}
\funstruct,  \cfg \vdash \seccontext \{\pc\} \seccontext[\sregr \mapsto \hseclabel, \srege \mapsto \lseclabel]
\end{array}
\endprooftree 
\end{math}
\\[20pt]

\begin{math}
\prooftree
\begin{array}{c}
C=\!\bnfc{icall}{e_f}\!\of{e_1,\!\dots,\!e_4}\ \ \seccontext \vdash e_f\!:\!\seclabel_f \sqsubseteq \lseclabel \ \ \seccontext \vdash e_i\!:\!\seclabel_i\ \textbf{for}\ i\!=\!1\!\dots\!4 \\[5pt]
\forall \cntrl \in \pred{\cfg}{\pc}.\ \elem{\cntrl}{C}\!=\! 
\bnfc{assert}\!\left(\!
\begin{array}{ll}
\secle{\cmagicn}{\seclabel_{[1-4]}} & \textbf{and} \\
e_f \in \cfg & \textbf{and} \\
({e_f}\!\mapsto\!\tuple{-,\cmagicn})\!\in\!\funstruct)  &
\end{array}\!\right)
\end{array}
\justifies
\begin{array}{l}
\funstruct,  \cfg \vdash \seccontext \{\pc\} \seccontext[\sregr \mapsto \hseclabel, \srege \mapsto \lseclabel]
\end{array}
\endprooftree 
\end{math}
\\[30pt]

\begin{math}
\prooftree
\begin{array}{c}
C = \bnfc{ret}\ \ \ \ \seccontext \vdash \srege\!\sqsubseteq\!\lseclabel \ \ \ \ \seccontext \vdash \reg{0}: \seclabel\\[5pt]
\forall\!\cntrl\!\in\!\pred{\cfg}{\pc}.\ \elem{\cntrl}{C}\!=\!\bnfc{assert}(\forall \cntrl'\!\in\!\pred{\cfg}{\mathit{top}(\stack_{\lseclabel})}.\ \secle{\cntrl'.\seccontext'(\reg{0})}{\seclabel})
\end{array}
\justifies
\begin{array}{l}
\funstruct,  \cfg \vdash \seccontext\ \{\pc\}\ \seccontext
\end{array}
\endprooftree 
\end{math}
\\[20pt]

\begin{math}
\prooftree
\begin{array}{l}
C = \bnfc{goto}\of{e}\ \ \ \  \seccontext \vdash e:\seclabel_e \sqsubseteq \lseclabel
\end{array}
\justifies
\begin{array}{l}
\funstruct, \cfg \vdash \seccontext\ \{\pc\}\ \seccontext
\end{array}
\endprooftree 
\end{math}
\\[20pt]

\begin{math}
\prooftree
\begin{array}{c}
C = \beifthenelse{e}{\bcjump{e_1}}{\bcjump{e_2}} \ \  \seccontext \vdash e: \seclabel_e  \sqsubseteq \lseclabel
\end{array}
\justifies
\begin{array}{l}
\funstruct, \cfg \vdash \seccontext\ \{\pc\}\ \seccontext
\end{array}
\endprooftree 
\end{math}

\end{tabular}
}
\caption{Complete list of type rules. $C$ is a command from the CFG node pointed to by $\pc$ and $\srege$ and $\sregr$ are callee- and caller-save registers. } 
\label{apx:symsem}
\end{figure}

\renewcommand{\cons}[1]{\underset{#1}{\mathit{Cons}}}

\end{document}